\documentclass[aps,prd,superscriptaddress,showkeys,nofootinbib]{revtex4-2}
\usepackage[english]{babel}
\usepackage{amsmath,bm}
\usepackage{amssymb}
\usepackage{enumitem}
\usepackage{textcomp}
\usepackage{graphicx}
\usepackage{dsfont}
\usepackage{color}

\definecolor{GREEN}{rgb}{0.,0.5,0}
\definecolor{BLUE}{rgb}{0.,0.,0.75}

\usepackage[breaklinks=true]{hyperref}
\hypersetup{
	colorlinks=true,
	linkcolor=BLUE,
	filecolor=magenta,      
	urlcolor=BLUE,
	citecolor=BLUE,
}

\usepackage{mathrsfs}
\usepackage[normalem]{ulem}

\usepackage{placeins}

\begin{document}

{\hfill CERN-TH-2021-128}
	
	\title{Gauge-field production during axion inflation\\in the gradient expansion formalism}
	
	\author{E.V.~Gorbar}
	\affiliation{Physics Faculty, Taras Shevchenko National University of Kyiv, 64/13, Volodymyrska Str., 01601 Kyiv, Ukraine}
	\affiliation{Bogolyubov Institute for Theoretical Physics, 14-b, Metrologichna Str., 03143 Kyiv, Ukraine}
	
	\author{K.~Schmitz}
	\affiliation{Theoretical Physics Department, CERN, 1211 Geneva 23, Switzerland}
	
	\author{O.O.~Sobol}
	\email{oleksandr.sobol@epfl.ch}
	\affiliation{Institute of Physics, Laboratory of Particle Physics and Cosmology, \'{E}cole Polytechnique F\'{e}d\'{e}rale de Lausanne, CH-1015 Lausanne, Switzerland}
	\affiliation{Physics Faculty, Taras Shevchenko National University of Kyiv, 64/13, Volodymyrska Str., 01601 Kyiv, Ukraine}
	
	\author{S.I.~Vilchinskii}
	\affiliation{Physics Faculty, Taras Shevchenko National University of Kyiv, 64/13, Volodymyrska Str., 01601 Kyiv, Ukraine}
	\affiliation{D\'{e}partement de Physique Th\'{e}orique, Center for Astroparticle Physics, Universit\'{e} de Gen\`{e}ve, 1211 Gen\`{e}ve 4,  Switzerland}
	
	\date{\today}
	
	\begin{abstract}
    We study the explosive production of gauge fields during axion inflation in a novel gradient expansion formalism that describes the time evolution of a set of bilinear electromagnetic functions in position space. Based on this formalism, we are able to simultaneously account for two important effects that have thus far been mostly treated in isolation: (i) the backreaction of the produced gauge fields on the evolution of the inflaton field and (ii) the Schwinger pair production of charged particles in the strong gauge-field background. This allows us to show that the suppression of the gauge-field production due to the Schwinger effect can prevent the backreaction in scenarios in which it would otherwise be relevant. Moreover, we point out that the induced current, $\bm{J} = \sigma \bm{E}$, also dampens the Bunch-Davies vacuum fluctuations deep inside the Hubble horizon. We describe this suppression by a new parameter $\Delta$ that is related to the time integral over the conductivity $\sigma$ which hence renders the description of the entire system inherently nonlocal in time. Finally, we demonstrate how our formalism can be used to construct highly accurate solutions for the mode functions of the gauge field in Fourier space, which serves as a starting point for a wealth of further phenomenological applications, including the phenomenology of primordial perturbations and baryogenesis.
	\end{abstract}

	\maketitle

	\section{Introduction}
	\label{sec-intro}
	
Gamma-ray observations from distant blazars provide indirect evidence for the presence of magnetic fields in voids of our Universe \cite{Tavecchio:2010,Ando:2010,Neronov:2010,Tavecchio:2011,Dolag:2010,Dermer:2011,Taylor:2011,Huan:2011,Vovk:2012,Caprini:2015} (for a recent review, see Ref.~\cite{Batista:2021}). Together with observations of the cosmic microwave background (CMB) \cite{Jedamzik:2000,Planck:2015-pmf,Sutton:2017,Jedamzik:2019}, this constrains the strength of the fields in the range $10^{-17}\,\mbox{G}\lesssim B_{0}\lesssim 10^{-10}\,$G. The very large coherence length of these fields \cite{Batista:2020}, measured in Megaparsecs, implies that they are of cosmological nature, because astrophysical mechanisms are inefficient in voids due to their small matter content. In principle, it is possible to inject magnetic fields into voids through outflows of matter from galaxies and active galactic nuclei \cite{Furlanetto:2001,Bertone:2006,Samui:2018,Garcia:2020}; however, it is quite problematic to attain an extremely large correlation length of the observed fields.
The possible cosmological origin of magnetic fields in voids has another important advantage because such primordial fields could serve as seed magnetic fields for galaxies and galaxy clusters \cite{Parker:1971}. These seed magnetic fields could be later enhanced astrophysically through different types of dynamo \cite{Zeldovich:1980book,Lesch:1995,Kulsrud:1997,Colgate:2001,Vazza:2018} and adiabatic compression \cite{Grasso:2001} to the strength of magnetic fields observed in galaxies and galaxy clusters \cite{Grasso:2001,Kronberg:1994,Widrow:2002,Giovannini:2004,Kandus:2011,Vallee:2011,Ryu:2012,Durrer:2013,Subramanian:2016}.

The mechanism of inflationary magnetogenesis proposed in Refs.~\cite{Turner:1988,Ratra:1992} naturally explains the large correlation length of magnetic fields in voids. This means that, in addition to the abundance of chemical elements, the CMB radiation, and the large-scale structure of galaxies and their clusters, the observed magnetic fields in voids could provide a unique source of information about physical processes in the early Universe at very high energies that are unattainable in the laboratory.
Since Maxwell's action is conformally invariant, fluctuations of the massless electromagnetic field are not enhanced in a conformally flat inflationary background \cite{Parker:1968}. Consequently, the conformal invariance should be necessarily broken in order to generate gauge fields during inflation. Such a breaking can be done in may ways, e.g., by introducing an interaction with a scalar or pseudoscalar inflaton field or with spacetime curvature \cite{Turner:1988,Ratra:1992,Garretson:1992,Dolgov:1993}. In our study, we will consider the axial coupling of an Abelian gauge field to a pseudoscalar ``axion'' inflaton field, which has attracted a lot of attention in the literature \cite{Anber:2006,Anber:2010,Durrer:2011,Barnaby:2012,Caprini:2014,Anber:2015,Ng:2015,Fujita:2015,Adshead:2015,Adshead:2016,Notari:2016,Jimenez:2017cdr,Domcke:2018,Cuissa:2018,Shtanov:2019,Shtanov:2019b,Sobol:2019,Domcke:2019bar,Domcke:2019,Domcke:2020,Kamarpour:2021a,Kamarpour:2021b}, mainly because it produces magnetic fields with nonzero helicity, which facilitates their survival in the primordial plasma after inflation.
 
In addition to magnetic fields, strong electric fields are always generated in inflationary models. Due to the Schwinger effect \cite{Sauter:1931,Heisenberg:1936,Schwinger:1951}, these strong electric fields produce pairs of charged particles and antiparticles and quickly form an ultrarelativistic plasma. Such a plasma efficiently screens electric fields and, therefore, strongly influences the generation and evolution of the gauge fields, especially near the end of inflation and during reheating. The Schwinger pair production during inflation and its impact on magnetogenesis was investigated in Refs.~\cite{Domcke:2018,Sobol:2019,Domcke:2019,Kobayashi:2014,Froeb:2014,Bavarsad:2016,Stahl:2016a,Stahl:2016b,Hayashinaka:2016a,Hayashinaka:2016b,Sharma:2017,Bavarsad:2018,Geng:2018,Hayashinaka:2018,Hayashinaka:thesis,Giovannini:2018a,Banyeres:2018,Stahl:2018,Kitamoto:2018,Sobol:2018,Shtanov:2020,Tangarife:2017,Chua:2019,Shakeri:2019,Gorbar:2019,Sobol:2020Sch}. For a recent application in the context of cosmological relaxation as a solution to the hierarchy problem, see also Ref.~\cite{Domcke:2021yuz}.

To handle the complicated dynamics of the inflaton and gauge fields, taking into account the Schwinger effect and the backreaction of the generated gauge fields and primordial plasma on the cosmological evolution, a novel gradient expansion formalism was developed by three of us in Ref.~\cite{Sobol:2019}. The formalism was later extended also to the case of the kinetic-coupling model \cite{Sobol:2020}. In the standard approach to magnetogenesis, one works with separate Fourier modes of the gauge field which evolve in a given inflationary background. In contrast, in the gradient expansion formalism, one considers vacuum expectation values of a truncated set of bilinear functions of the electric and magnetic fields in coordinate space that include all physically relevant modes at once. It was shown that, even taking into account a relatively small number of bilinear functions, it is possible to describe the electric and magnetic energy densities with an accuracy of a few percent during the whole stage of inflation. The formalism also takes into account the fact that the number of relevant modes constantly grows during inflation as new modes cross the horizon and undergo the quantum to classical transition.\footnote{Here and in the following, ``horizon crossing'' will denote the time when a given mode with momentum $k$ becomes tachyonically unstable, which typically happens slightly before it leaves the actual Hubble horizon, when $k= aH$, with scale factor $a$ and Hubble rate $H$.} To account for the growth of the number of relevant modes outside the horizon, boundary terms are added to the equations of motion for the electromagnetic bilinear functions.

In the present paper, we will study the problem of generating helical magnetic fields in a model with an axial coupling of the gauge field to the inflaton. Using the gradient expansion formalism of Ref.~\cite{Sobol:2019}, which takes into account the backreaction of the generated fields and the Schwinger effect, we will develop an approximation scheme that allows us to reach an unprecedented accuracy in the numerical results\,---\,the error compared to the corresponding mode by mode solution is always less than one to two percent during inflation\,---\,and compare our results to those in the literature. Here, a characteristic feature and advantage of our formalism consists in the fact that we do not rely on an iterative procedure that needs to be repeated over and over again before it converges to a self-consistent result. Instead, our evolution equations only need to be evolved forward in time once in order to generate the desired output. 

We will be particularly interested in the case in which the Abelian gauge field coupling to the axion inflaton field is identified with the hypercharge gauge field in the Standard Model. Consequently, we will evaluate the electric conductivity $\sigma$ of the charged plasma generated during inflation based on the Standard Model particle content, assuming that all fermions remain massless during inflation. Specifically, we will use the expressions for the induced current and electric conductivity in Ref.~\cite{Domcke:2019}, which account for the presence of both electric and magnetic background fields. On the technical side, we will work out an improved description of the vacuum fluctuations deep inside the horizon that reflects the damping of the gauge-field amplitude by the charged plasma. This will lead us on the one hand to improved boundary terms in our evolution equations and on the other hand to the notion of a damped Bunch-Davies vacuum, which we describe in terms of a new parameter $\Delta$. As we will see, $\Delta$ is related to the time integral over the conductivity $\sigma$ and hence is nonlocal in time. An important outcome of our analysis therefore is the fact that the state of the system at any given moment in time $t$ does not only depend on the choice of parameter values in the Lagrangian and quantities such as the instantaneous velocity of the inflaton field; it also depends on the entire prehistory leading up to this moment in time, which controls to what extent the vacuum fluctuations of modes inside the horizon have already been damped in the course of inflation at times $t' \leq t$.

The paper is organized as follows. The axial-coupling model for generating magnetic fields during inflation is described in Sec.~\ref{sec-model}. In Sec.~\ref{sec-GEF}, the system of equations for the electromagnetic bilinear quantities is derived using the gradient expansion formalism. Numerical solutions of this system of equations are given and discussed in Sec.~\ref{sec-numerical}. In this section, we will compare the outcome of our analysis to other methods and results available in the literature whenever possible. This will allow us to validate our formalism; first in the case of no backreaction and no Schwinger effect and then subsequently in the case of relevant backreaction and no Schwinger effect. In the third step, we will then extend our analysis and include the Schwinger effect, which goes beyond existing results in the literature. An important result of this part of our analysis will be that Schwinger pair production can suppress the energy density of the electromagnetic field to such an extent that no backreaction effects occur after all, even though backreaction effects would be absolutely essential in the absence of the Schwinger pair production. In addition, we will present the Fourier spectra of the electric and magnetic fields in this section as well as their relation to the spectra in the (damped) Bunch-Davies vacuum. Section~\ref{sec-concl} finally summarizes the findings obtained in this paper and contains an outlook on possible future steps. In this section, we will also comment on possible phenomenological applications of our results, such as the generation of primordial scalar and tensor perturbations and baryogenesis from hypermagnetic fields. Throughout the paper we use natural units and set $\hbar=c=1$ with the reduced Planck mass equal to $M_{\mathrm{P}}=(8\pi G)^{-1/2}=2.435\times 10^{18}\,$GeV. We assume that the Universe is described on cosmological scales by a spatially flat Friedmann-Lema\^{i}tre-Robertson-Walker (FLRW) metric in terms of the cosmic time, $g_{\mu\nu}={\rm diag\,}\{1, \,-a^{2}(t),\, -a^{2}(t),\, -a^{2}(t)\}$.

\section{Axial-coupling model}
\label{sec-model}

Let us describe the axial-coupling model that we use for the analysis of the generation of primordial magnetic fields. In this model, the conformal invariance of the Maxwell action is broken by means of the axial coupling of the electromagnetic field $A_{\mu}$ with the pseudoscalar inflaton field $\phi$. The corresponding action reads as\footnote{In fact, $A_{\mu}$ represents a generic Abelian gauge field axially coupled to the inflaton.  However, for simplicity, we will sometimes refer to it as the electromagnetic field. Our numerical analysis will be performed for the specific case of the Standard Model hypercharge field.}	
\begin{equation}
\label{S}
S=\int d^{4}x\sqrt{-g} \left[\frac{1}{2}g^{\mu\nu}\,\partial_{\mu}\phi\,\partial_{\nu}\phi - V(\phi)-\frac{1}{4}F_{\mu\nu}F^{\mu\nu}-\frac{1}{4}I(\phi)F_{\mu\nu}\tilde{F}^{\mu\nu} +\mathcal{L}_{\rm ch}(\chi,\,A_{\mu})\right],
\end{equation}
where $g={\rm det\,}g_{\mu\nu}$ is the determinant of the spacetime metric, $V(\phi)$ is the inflaton potential, $I(\phi)$ is the axial-coupling function, $F_{\mu\nu}=\partial_{\mu}A_{\nu}-\partial_{\nu}A_{\mu}$ is the gauge-field strength tensor, and	
\begin{equation}
\label{F}
\tilde{F}^{\mu\nu}=\frac{1}{2\sqrt{-g}}\,\varepsilon^{\mu\nu\lambda\rho}F_{\lambda\rho}
\end{equation}
is the corresponding dual tensor; $\varepsilon^{\mu\nu\lambda\rho}$ is the absolutely antisymmetric Levi-Civita symbol with $\varepsilon^{0123}=+1$. The last term in Eq.~(\ref{S}) is the gauge-invariant Lagrangian of a generic matter field $\chi$ charged under the $U(1)$ gauge group and, therefore, coupled to the electromagnetic four-potential $A_{\mu}$. For the sake of generality, we will not specify this term and assume that it describes all charged fields in the model. 
	
Action (\ref{S}) implies the following Euler-Lagrange equations for the inflaton and electromagnetic fields,
\begin{equation}
\label{L_E_1}
\frac{1}{\sqrt{-g}}\partial_{\mu} \left[\sqrt{-g}\, g^{\mu\nu}\,\partial_{\nu}\phi \right] + \frac{dV}{d\phi} + \frac{1}{4}\frac{dI}{d\phi}F_{\mu\nu}\tilde{F}^{\mu\nu}=0,
\end{equation}
\begin{equation}
\label{L_E_2}
\frac{1}{\sqrt{-g}}\partial_{\mu}\left[\sqrt{-g}\,F^{\mu\nu} \right]+ \frac{dI}{d\phi}\,\tilde{F}^{\mu\nu}\partial_{\mu}\phi=j^{\nu},
\end{equation}
where 
\begin{equation}
    j^{\nu}=-\frac{\partial \mathcal{L}_{\rm ch}(\chi,\,A_{\mu})}{\partial A_{\nu}}
\end{equation}
is the electric four-current. Equation (\ref{L_E_2}) should be supplemented by the Bianchi identity for the dual gauge field strength tensor
\begin{equation}
\label{Bianchi}
\frac{1}{\sqrt{-g}}\partial_{\mu}\left[\sqrt{-g}\,\tilde{F}^{\mu\nu} \right] = 0.
\end{equation}
The energy-momentum tensor equals
\begin{equation}
\label{T}
T_{\mu\nu}=\frac{2}{\sqrt{-g}}\,\frac{\delta S}{\delta g^{\mu\nu}}=\partial_{\mu}\phi\,\partial_{\nu}\phi - g^{\lambda\rho}F_{\mu\lambda}F_{\nu\rho}-g_{\mu\nu}\left[\frac{1}{2}\partial_{\alpha}\phi\,\partial^{\alpha}\phi-V(\phi)-\frac{1}{4}F_{\alpha\beta}F^{\alpha\beta}\right]+T_{\mu\nu}^{\chi},
\end{equation}
where the last term describes the contribution of the charged matter fields. 

Now we use the FLRW metric and choose the Coulomb gauge for the vector potential $A_{\mu}$ where $A_{\mu}=(0,\,-\boldsymbol{A})$. Then, the three-vectors of electric $\bm{E}=(E^{1},\,E^{2},\,E^{3})$ and magnetic $\bm{B}=(B^{1},\,B^{2},\,B^{3})$ fields can be defined as $\bm{E}=-\tfrac{1}{a}\partial_{0}\bm{A}$ and $\bm{B}=\tfrac{1}{a^{2}}{\rm rot\,}\bm{A}$. Note that these are the physically measured fields by a comoving observer; therefore, we included the scale factor in their definition. The gauge field stress tensor and its dual tensor are expressed in terms of the electric and magnetic fields as follows:
\begin{equation}
F_{0i}=aE^{i}, \quad F_{ij}=-a^{2}\varepsilon_{ijk} B^{k}, \quad \tilde{F}_{0i}=aB^{i}, \quad \tilde{F}_{ij}=a^{2}\varepsilon_{ijk} E^{k}.
\end{equation}
The cosmic expansion rate (the Hubble parameter $H=\dot{a}/a$) is determined by the Friedmann equation
\begin{equation}
\label{Friedmann}
H^{2}=\frac{\rho}{3 M_{\mathrm{P}}^{2}},
\end{equation}
where the total energy density $\rho$ is given by the zero-zero component of the energy-momentum tensor (\ref{T}),
\begin{equation}
\label{rho}
\rho=T_{0}^{0}=\left[\frac{1}{2}\dot{\phi}^{2}+V(\phi) \right] + \frac{1}{2}\left\langle \bm{E}^{2}+\bm{B}^{2} \right\rangle +\rho_{\chi}.
\end{equation}
Here the two terms in square brackets correspond to the spatially homogeneous inflaton field, the next term gives the gauge-field contribution (angular brackets denote the vacuum expectation value), while the last one is the counterpart for the charged matter fields. 

The electric current four-vector can be decomposed as
\begin{equation}
    j^{\mu}=\big(\rho_{\rm c},\,\frac{1}{a}\bm{J}\big).
\end{equation}
We assume that charged particles were absent in the Universe initially and were produced only later in particle-antiparticle pairs. Therefore, we set the charge density to zero, $\rho_{\rm c}=0$. On the other hand, the current density three-vector $\bm{J}$ may be nonzero in the presence of the electromagnetic field. Then the equations of motion (\ref{L_E_1})--(\ref{L_E_2}) and (\ref{Bianchi}) take the following form in the three-vector notation,
\begin{equation}
\label{KGF}
\ddot{\phi}+3H\dot{\phi}+V'(\phi)=I'(\phi)\left\langle \bm{E}\cdot\bm{B} \right\rangle,
\end{equation}
\begin{equation}
\label{Maxwell_1}
\dot{\bm{E}}+2 H \bm{E}-\frac{1}{a} {\rm rot} \bm{B} + I'(\phi)\,\dot{\phi}\,\bm{B}+\bm{J}=0,
\end{equation}
\begin{equation}
\label{Maxwell_2}
\dot{\bm{B}}+2 H \bm{B}+\frac{1}{a} {\rm rot} \bm{E}=0,
\end{equation}
\begin{equation}
\label{Maxwell_3}
{\rm div} \bm{E}=0, \qquad {\rm div} \bm{B}=0.
\end{equation}
Finally, in order to close the system of Maxwell's equations, we need to specify the electric current. It was shown in the literature (see, e.g., Refs.~\cite{Kobayashi:2014,Hayashinaka:2016a,Domcke:2019}) that the induced current of charged particles produced by the Schwinger effect in a constant electromagnetic field in de Sitter spacetime satisfies Ohm's law,
\begin{equation}
\label{Ohms-law}
    \bm{J}=\sigma\bm{E},
\end{equation}
where $\sigma$ is the generalized conductivity, which depends only on the absolute values of electric and magnetic fields. We will utilize this approximation in our analysis too, even though we will consider time-dependent electric and magnetic fields. The applicability of this approximation will be discussed in detail in Sec.~\ref{subsec-Schwinger}.

\section{Gradient expansion formalism}
\label{sec-GEF}

In this section, we derive the system of equations describing the self-consistent evolution of the inflaton, gauge fields, and charged particles produced by the Schwinger effect during axion inflation.

\subsection{Equations for bilinear electromagnetic functions}
	
Solving the system of Maxwell's equations requires determining the dependence of the gauge field on the spatial coordinates, which makes numerical calculations very demanding. Therefore, as mentioned in the Introduction, it is advantageous to apply the gradient expansion formalism, where one considers vacuum expectation values of bilinear electromagnetic functions in coordinate space that include all physically relevant modes at once. These functions are
\begin{equation}
\label{E_n}
\mathcal{E}^{(n)}=\frac{1}{a^{n}}\left\langle \bm{E}\cdot {\rm rot}^{n} \bm{E}  \right\rangle,
\end{equation}
\begin{equation}
\label{G_n}
\mathcal{G}^{(n)}=-\frac{1}{a^{n}}\left\langle \bm{E}\cdot {\rm rot}^{n} \bm{B}  \right\rangle,
\end{equation}	
\begin{equation}
\label{B_n}
\mathcal{B}^{(n)}=\frac{1}{a^{n}}\left\langle \bm{B}\cdot {\rm rot}^{n} \bm{B}  \right\rangle
\end{equation}
and contain spatial derivatives only as powers of the curl. This allows us to get a closed system of equations for these quantities, because  Maxwell's equations (\ref{Maxwell_1}) and (\ref{Maxwell_2}) contain spatial derivatives only in such a form. Using Eqs.(\ref{Maxwell_1})--(\ref{Maxwell_3}), we get the following equations of motion for the bilinear electromagnetic functions,
\begin{equation}
\label{dot_E_n}
\dot{\mathcal{E}}^{(n)} + [(n+4)H+2\sigma]\,	\mathcal{E}^{(n)} - 2I'(\phi)\dot{\phi}\,\mathcal{G}^{(n)} +2\mathcal{G}^{(n+1)}=[\dot{\mathcal{E}}^{(n)}]_{b},
\end{equation}
\begin{equation}
\label{dot_G_n}
\dot{\mathcal{G}}^{(n)} +[(n+4)H+\sigma]\, \mathcal{G}^{(n)}-\mathcal{E}^{(n+1)}+\mathcal{B}^{(n+1)} - I'(\phi)\dot{\phi}\,\mathcal{B}^{(n)}=[\dot{\mathcal{G}}^{(n)}]_{b},
\end{equation}
\begin{equation}
\label{dot_B_n}
\dot{\mathcal{B}}^{(n)} + (n+4)H\,	\mathcal{B}^{(n)}-2\mathcal{G}^{(n+1)}=[\dot{\mathcal{B}}^{(n)}]_{b}.
\end{equation}
Note that the equation of motion for the $n$th order function always contains at least one function with the $(n+1)$th power of the curl. As a result, all equations are coupled into an infinite chain that needs to be truncated in practice.

The right-hand sides of Eqs.(\ref{dot_E_n})--(\ref{dot_B_n}) contain the contributions due to boundary terms. Their necessity is dictated by the following consideration. Modes inside the horizon correspond to vacuum fluctuations of the quantum gauge field. They oscillate in time without significant change of their amplitude. Therefore, their contribution to the electromagnetic energy density should be excluded. However, modes outside the horizon do not oscillate and can be treated as the Fourier modes of a classical electromagnetic field \cite{Lyth:2008}. During inflation, modes that were initially inside the horizon cross the horizon and begin to behave classically. Therefore, the number of physically relevant modes (outside the horizon) constantly grows during inflation. To take this fact into account, one needs to introduce the corresponding boundary terms for the bilinear electromagnetic functions. The resulting system of equations forms a closed set of ordinary differential equations that describe the self-consistent evolution of classical observables in the form of quadratic functions of the electric and magnetic fields with an arbitrary power of the curl.

We would like to mention that the boundary terms for the system of Eqs.~(\ref{dot_E_n})--(\ref{dot_B_n}) were derived for the first time in Ref.~\cite{Sobol:2019}. In this study, we calculate the boundary terms more accurately, including the impact of the Schwinger effect. In addition, we propose a new way of truncating the infinite chain of equations of the system that allows us to reach a high accuracy in the numerical results.

\subsection{Schwinger conductivity}
\label{subsec-Schwinger}

Before calculating the boundary terms, let us discuss the Schwinger pair production. In constant and uniform collinear electric and magnetic fields in an expanding Universe, this effect was studied in Ref.~\cite{Domcke:2019}. In the case of one Dirac fermion species with mass $m$ and charge $Q$, the conductivity induced by the Schwinger effect has the form
\begin{equation}
\label{sigma}
    \sigma_{\mathrm{f}}=\frac{(e|Q|)^{3}}{6\pi^{2}}\frac{|B|}{H}{\rm coth}\Big(\frac{\pi |B|}{|E|}\Big)\exp\Big(-\frac{\pi m^{2}}{|eQE|} \Big),
\end{equation}
where $|B|$ and $|E|$ are the absolute values of the collinear magnetic and electric fields and $e$ is the $U(1)$ gauge coupling constant. The contribution of a single Weyl fermion is twice smaller. It is important to remember that this expression was derived in the strong-field regime, $|eE|\gg H^{2}$, which is the most important one for physical applications. This expression is in good agreement with the result for flat Minkowski spacetime (see, e.g., Refs.~\cite{Nikishov:1969,Dunne:2004,Ruffini:2009,Warringa:2012}) if we replace $1/(3H)$ with the time $\Delta t$ during which the external field is switched on. Moreover, for small magnetic field $|B|\ll |E|$, it reduces to the expression for the fermionic Schwinger conductivity in a purely electric field in de Sitter spacetime, $\sigma_{\mathrm{f},0}=(e|Q|)^{3}|E|/(6\pi^{3}H)\exp(-\pi m^{2}/|eQE|)$ obtained in Ref.~\cite{Hayashinaka:2016a}.

Using the same logic, it is easy to construct the Schwinger conductivity for scalar charge carriers in the strong-field regime. In flat Minkowski spacetime, the pair production rate for scalars differs from that of fermions with the replacement of ${\rm coth}(\pi |B|/|E|)$ by $(1/2){\rm cosech}(\pi|B|/|E|)$, where ${\rm cosech\,}x=1/{\rm sinh\,}x$ \cite{Dunne:2004,Ruffini:2009}. Then, in de Sitter spacetime, we expect that
\begin{equation}
\label{sigma-scalar}
    \sigma_{\mathrm{s}}=\frac{(e|Q|)^{3}}{12\pi^{2}}\frac{|B|}{H}{\rm cosech}\Big(\frac{\pi |B|}{|E|}\Big)\exp\Big(-\frac{\pi m^{2}}{|eQE|} \Big).
\end{equation}
A consistency of this result can be checked by taking the limit of a vanishingly small magnetic field, $|B|\ll |E|$. In this case, we obtain $\sigma_{\mathrm{s},0}=(e|Q|)^{3}|E|/(12\pi^{3}H)\exp(-\pi m^{2}/|eQE|)$, which is the well-known expression for the Schwinger conductivity in a constant electric field in de Sitter spacetime computed in Ref.~\cite{Kobayashi:2014}.

We would like to emphasize that expressions (\ref{sigma}) and (\ref{sigma-scalar}) for the Schwinger conductivity have been derived in the approximation of constant collinear electric and magnetic fields in the de Sitter background. Therefore, we assume that the electric and magnetic fields as well as the Hubble parameter evolve sufficiently slowly during inflation so that at any given moment of time the conductivity is determined by the corresponding values of these quantities.

Another our assumption is the collinearity of the electric and magnetic fields. In the axial-coupling model, the generated fields are almost maximally helical (for sufficiently large axial coupling). In such a case, the electric and magnetic fields are indeed nearly collinear. For instance, in the absence of the Schwinger effect and for $|\xi|\equiv |(dI/d\phi)(d\phi/dt)|/(2H)\gtrsim 1$, there exist quite simple analytical expressions for the generated fields \cite{Anber:2010,Sobol:2019},
\begin{equation}
    \langle\bm{E}^{2}\rangle \simeq \frac{9}{1120\pi^{3}}\frac{e^{2\pi|\xi|}H^{4}}{|\xi|^{3}}, \quad
    |\langle\bm{E}\cdot\bm{B}\rangle| \simeq \frac{9}{1120\pi^{3}}\frac{e^{2\pi|\xi|}H^{4}}{|\xi|^{4}}, \quad
    \langle\bm{B}^{2}\rangle \simeq \frac{1}{112\pi^{3}}\frac{e^{2\pi|\xi|}H^{4}}{|\xi|^{5}}. 
\end{equation}
Then the angle between the electric and magnetic fields can be estimated as
\begin{equation}
\label{estimate-theta}
    \theta=\arccos\frac{|\langle\bm{E}\cdot\bm{B}\rangle|}{\sqrt{\langle\bm{E}^{2}\rangle\langle\bm{B}^{2}\rangle}}\simeq \arccos \frac{3}{\sqrt{10}}\approx \frac{\pi}{10},
\end{equation}
which is indeed a relatively small angle. This is the minimal possible value of $\theta$ which can be reached in the free case without backreaction and the Schwinger effect in the limit $|\xi|\gg 1$. For smaller $|\xi|$, the angle can be greater, while in the presence of the Schwinger effect it can be even smaller than the value in Eq.~(\ref{estimate-theta}), see Figs.~\ref{fig-b10-SE}(b) and \ref{fig-b25-SE}(b) in Sec.~\ref{sec-numerical}.

In the subsequent analysis, we will assume that the $U(1)$ gauge group is the Standard Model hypercharge group $U(1)_{Y}$ and $e=g'$, $Q=Y$. We consider the electroweak-symmetric phase where all SM fermions are massless. This requires the SM Higgs field to be stabilized at the origin in field space during inflation, since any nonzero Higgs field value would otherwise break the electroweak symmetry and render the SM fermions massive (except for neutrinos, possibly). One possibility to stabilize the Higgs field consists, e.g., in a nonminimal coupling to the Ricci curvature scalar $R$, which endows the Higgs field with an effective mass of the order of $|R| \approx 12 H^2$ times a positive coupling constant during inflation. In the following, we will therefore focus on scenarios in which the Higgs field is very heavy during inflation, $m_{\mathrm{H}}^{2}\gg g'|E|$, such that its contribution to the Schwinger current can be neglected. We express the values of the electric and magnetic fields in the conductivity through the bilinear functions of the gauge field, i.e., we use $|B|=\sqrt{\mathcal{B}^{(0)}}$ and $|E|=\sqrt{\mathcal{E}^{(0)}}$, and finally get the following result (see Appendix~\ref{app-particles} for more details),
\begin{equation}
\label{sigma-SM}
    \sigma_{\mathrm{SM}}=
    \frac{41g^{\prime 3}}{72\pi^{2}}\frac{\sqrt{\mathcal{B}^{(0)}}}{H}{\rm coth}\Big(\pi\sqrt{\frac{\mathcal{B}^{(0)}}{\mathcal{E}^{(0)}}}\Big).
\end{equation}

We take into account the running of the coupling constant $g'$ due to all SM particles \cite{Srednicki-book},
\begin{equation}
\label{running}
    [g^{\prime}(\mu)]^{-2}=
    [g^{\prime}(m_{Z})]^{-2}+\frac{41}{48\pi^{2}}\ln\frac{m_{Z}}{\mu}.
\end{equation}
At the energy scale of the $Z$-boson mass, $m_{Z}\approx 91.2\,$GeV, the gauge coupling equals $g^{\prime}(m_{Z})\approx 0.35$. As a characteristic energy scale $\mu$ relevant for the Schwinger pair production, we use
\begin{equation}
    \mu=\rho_{\rm em}^{1/4}=\Big(\frac{\langle\bm{E}^{2}\rangle+\langle\bm{B}^{2}\rangle}{2}\Big)^{1/4}.
\end{equation}
The coefficient in front of the logarithmic term in Eq.~(\ref{running}) may be affected by mass threshold effects related to the mass of the Higgs field during inflation. However, the Higgs contribution is 40 times smaller than that of all SM fermions, so these effects can only lead to very small changes, therefore, we will neglect them in the following.

Finally, we would like to discuss the evolution of the energy density of the produced charged particles, $\rho_{\chi}$, which affects the cosmic expansion rate. The equation of motion for this quantity can be derived from the energy conservation law. Indeed, multiplying Eq.~(\ref{KGF}) for the inflaton field by $\dot{\phi}$, we rewrite it in the form
\begin{equation}
\label{dot-rho-inf}
    \dot{\rho}_{\rm inf}+3H (\rho_{\rm inf}+p_{\rm inf})=-I'(\phi)\dot{\phi}\mathcal{G}^{(0)},
\end{equation}
where $\rho_{\rm inf}=\dot{\phi}^{2}/2+V(\phi)$ and $p_{\rm inf}=\dot{\phi}^{2}/2-V(\phi)$ are the energy density and the pressure of the inflaton field, respectively. Further, the equation describing the evolution of the electromagnetic energy density $\rho_{\rm em}=\langle \bm{E}^{2}+\bm{B}^{2}\rangle/2=(\mathcal{E}^{(0)}+\mathcal{B}^{(0)})/2$ can be obtained from Eqs.~(\ref{dot_E_n}) and (\ref{dot_B_n}) for $n=0$,
\begin{equation}
\label{dot-rho-EB}
    \dot{\rho}_{\rm em}+4H\rho_{\rm em}=[\dot{\rho}_{\rm em}]_{b}+I'(\phi)\dot{\phi}\mathcal{G}^{(0)}-\sigma \mathcal{E}^{(0)}.
\end{equation}
The first term on the right-hand side describes the increase of the electromagnetic energy density due to new modes crossing the horizon during inflation; therefore, it can be thought of as a vacuum source term. Comparing Eqs.~(\ref{dot-rho-inf}) and (\ref{dot-rho-EB}), we conclude that the second term on the right-hand side of Eq.~(\ref{dot-rho-EB}) describes the energy transfer from the inflaton to the gauge field due to the axial coupling. Finally, the last term in this equation determines the energy loss due to the Schwinger effect. Then we have the following equation for the energy density of produced particles,
\begin{equation}
    \label{eq-rho-chi}
    \dot{\rho}_{\chi}+4H\rho_{\chi}=\sigma \mathcal{E}^{(0)},
\end{equation}
where we assumed that $m^{2}\ll g'|\bm{E}|$ so that the produced particles are ultrarelativistic; in fact, we assume them to be massless. In the opposite case, the factor $4H$ should be replaced by $3H$ as for nonrelativistic particles.

\subsection{Boundary terms}

In order to derive the explicit form of the boundary terms, let us consider the quantized gauge field
\begin{equation}
\label{quantized_A}
\hat{\bm{A}}(t,\bm{x})=\int\frac{d^{3}\bm{k}}{(2\pi)^{3/2}}\sum_{\lambda=\pm}\left[\boldsymbol{\epsilon}^{\lambda}(\bm{k})\hat{a}_{\bm{k},\lambda}A_{\lambda}(t,k)e^{i\bm{k}\cdot\bm{x}}+\boldsymbol{\epsilon}^{*\lambda}(\bm{k})\hat{a}_{\bm{k},\lambda}^{\dagger}A_{\lambda}^{*}(t,k)e^{-i\bm{k}\cdot\bm{x}} \right],
\end{equation}
where $A_{\lambda}(t,k)$ is the mode function, $\boldsymbol{\epsilon}^{\lambda}(\bm{k})$ is the polarization three-vector, $\hat{a}_{\bm{k},\lambda}$ ($\hat{a}^{\dagger}_{\bm{k},\lambda}$) is the annihilation (creation) operator of the electromagnetic mode with momentum $\bm{k}$ and circular polarization $\lambda=\pm$, and $k=|\bm{k}|$.
The polarization vectors satisfy the following properties
\begin{equation}
\bm{k}\cdot\boldsymbol{\epsilon}^{\lambda}(\bm{k})=0,\quad \boldsymbol{\epsilon}^{*\lambda}(\bm{k})=\boldsymbol{\epsilon}^{-\lambda}(\bm{k}), \quad [i\bm{k}\times\boldsymbol{\epsilon}^{\lambda}(\bm{k})]=\lambda k \boldsymbol{\epsilon}^{\lambda}(\bm{k}), \quad \boldsymbol{\epsilon}^{*\lambda}(\bm{k})\cdot\boldsymbol{\epsilon}^{\lambda'}(\bm{k})=\delta^{\lambda\lambda'}.
\end{equation}
The creation and annihilation operators have the canonical commutation relations
\begin{equation}
[\hat{a}_{\bm{k},\lambda},\,\hat{a}^{\dagger}_{\bm{k}',\lambda'}]=\delta_{\lambda\lambda'}\delta^{(3)}(\bm{k}-\bm{k}').
\end{equation}

Substituting decomposition (\ref{quantized_A}) into Eqs.~(\ref{E_n})--(\ref{B_n}) and taking the vacuum expectation values, we obtain
\begin{equation}
\label{E_1}
\mathcal{E}^{(n)}=\sum_{\lambda=\pm 1}\int_{0}^{k_{\mathrm{h}}}\frac{d k}{k} \lambda^{n}\frac{k^{n+3}}{2\pi^{2}a^{n+2}}\,|\dot{A}_{\lambda}(t,k)|^{2}=\sum_{\lambda=\pm 1}\int_{0}^{k_{\mathrm{h}}}\frac{d k}{k} \lambda^{n}\frac{k^{n+5}}{2\pi^{2}a^{n+4}}\,|D_{\lambda}(t,k)|^{2},
\end{equation}
\begin{equation}
\label{G_1}
\mathcal{G}^{(n)}=\sum_{\lambda=\pm 1}^{}\int_{0}^{k_{\mathrm{h}}}\frac{d k}{k} \lambda^{n+1}\frac{k^{n+4}}{4\pi^{2}a^{n+3}}\,\frac{d}{d t}|{A}_{\lambda}(t,k)|^{2}=\sum_{\lambda=\pm 1}^{}\int_{0}^{k_{\mathrm{h}}}\frac{d k}{k} \lambda^{n+1}\frac{k^{n+5}}{2\pi^{2}a^{n+4}}\Re e\left[A_{\lambda}^{*}D_{\lambda} \right],
\end{equation}
\begin{equation}
\label{B_1}
\mathcal{B}^{(n)}=\sum_{\lambda=\pm 1}^{}\int_{0}^{k_{\mathrm{h}}}\frac{d k}{k} \lambda^{n}\frac{k^{n+5}}{2\pi^{2}a^{n+4}}\,|A_{\lambda}(t,k)|^{2},
\end{equation}
where
\begin{equation}
\label{D}
D_{\lambda}(t,k)=\frac{a}{k}\frac{\partial A_{\lambda}}{\partial t}=\frac{1}{k}\frac{\partial A_{\lambda}}{\partial\eta}.
\end{equation}
Note that the integration in the above expressions proceeds over wave numbers less than $k_{\mathrm{h}}(t)$, which is the wave number of the mode that crosses the horizon at a given moment of time $t$ and whose explicit form will be given below.
Clearly, the dependence of $k_{\mathrm{h}}$ on time is the underlying reason for the appearance of the boundary terms.

Let us forget for a while that the conductivity depends on the electric and magnetic fields and consider $\sigma(t)$ simply as a function of time. Then, Eqs.~(\ref{L_E_2}) and (\ref{quantized_A}) tell us that the equation of motion for $A_{\lambda}(t,k)$ is given by
\begin{equation}
\label{A_1}
\ddot{A}_{\lambda}(t,k)+(H+\sigma)\dot{A}_{\lambda}(t, k)+\left[\frac{k^{2}}{a^{2}}-\lambda\frac{k}{a}\frac{dI}{d\phi}\dot{\phi}\right] A_{\lambda}(t,k) =0,
\end{equation}
where $\sigma(t)$ appears in the frictionlike term $\sigma \dot{A}_{\lambda}(t,k)$ on the left-hand side in addition to the usual Hubble friction term $H \dot{A}_{\lambda}(t,k)$. As we will see shortly, it is this additional friction term induced by the Schwinger conductivity that will lead us to the notion of a damped Bunch-Davies vacuum deep inside the horizon. The longer the friction term is active, the stronger will be the damping of the electromagnetic field. In order to find an approximate solution for the mode equation (\ref{A_1}), it is convenient to do the following changes of function and variable
\begin{equation}
\label{damping-mode-function}
A_{\lambda}(t,k)=\exp\Big(-\frac{1}{2}\int\limits_{-\infty}^{t}\sigma(t')dt'\Big)f_{\lambda}(z,k), \qquad z=k\eta(t),
\end{equation}
where $\eta(t)=\int^{t}dt'/a(t')$ is the conformal time. If we assume that during inflation the cosmic expansion can be approximated by the de Sitter solution $\eta\simeq -1/(aH)$, the function $f_{\lambda}$ satisfies the following equation
\begin{equation}
\label{A_2}
\frac{\partial^{2}f_{\lambda}(z,k)}{dz^{2}}+\left[1+\frac{\lambda I'_{\phi}\dot{\phi}}{Hz}-\frac{\sigma^{2}+2\sigma H+2\dot{\sigma}}{4H^{2}z^{2}}\right] f_{\lambda}(z, k)=0.
\end{equation}
For further convenience, we introduce the notations
\begin{equation}
\label{xi-s}
\xi(t)=\frac{dI}{d\phi}\frac{\dot{\phi}}{2H}, \qquad s(t)=\frac{\sigma(t)}{2H}.
\end{equation}

Although the Schwinger conductivity depends on electric and magnetic fields, we assume that they vary slowly during inflation. Therefore, the term with $\dot{\sigma}$ in square brackets of Eq.~(\ref{A_2}) can be neglected. Then we get the following simplified equation
\begin{equation}
\label{A_3}
\frac{\partial^{2}f_{\lambda}(z,k)}{dz^{2}}+\left[1+\frac{2\lambda \xi}{z}-\frac{s^{2}+s}{z^{2}}\right] f_{\lambda}(z, k)=0.
\end{equation}
Deep inside the horizon, when the first term in square brackets dominates, the solution must satisfy the Bunch-Davies boundary condition \cite{Bunch:1978}
\begin{equation}
\label{BD}
f_{\lambda}(z, k)=\frac{1}{\sqrt{2k}}e^{-iz}, \quad -z\gg 1.
\end{equation}
We emphasize that Eq.~(\ref{BD}) does not fully describe the gauge-field mode function inside the horizon in the presence of finite conductivity. Indeed, the full mode function is given by Eq.~(\ref{damping-mode-function}) and includes an exponential damping factor,
\begin{equation}
\label{BD-damped}
A_{\lambda}(z,k)=\sqrt{\frac{\Delta}{2k}}e^{-iz}, \quad -z\gg 1,
\end{equation}
where we introduced the new parameter $\Delta$, which is defined in terms of the exponential of the integrated conductivity,
\begin{equation}
\label{Delta-parameter}
\Delta(t)\equiv \exp\Bigg(-\int\limits_{-\infty}^{t}\sigma(t')dt'\Bigg).
\end{equation}
This parameter suppresses the gauge-field amplitude on small scales. It is, moreover, nonlocal in time, since it depends on the conductivity at all times $t'\leq t$.
	
A mode crosses the horizon when the expression in the square brackets in Eq.~(\ref{A_3}) vanishes for the first time at least for one polarization, which is the case when
\begin{equation}
\label{horizon-crossing-time}
\frac{k}{a(t)H(t)}=|\xi(t)|+\sqrt{\xi^{2}(t)+s^{2}(t)+s(t)} \quad \Rightarrow \quad t=t_{\mathrm{h}}(k).
\end{equation}
In the vicinity of the horizon crossing when $t\approx t_{\mathrm{h}}$, $\xi(t)$ and $s(t)$ in Eq.~(\ref{A_3}) can be replaced by $\xi(t_{\mathrm{h}})={\rm constant}$ and $s(t_{\mathrm{h}})={\rm constant}$, respectively. Then the solution of Eq.~(\ref{A_3}) satisfying the Bunch-Davies boundary condition (\ref{BD}) can be expressed in terms of the Whittaker function (see Appendix~\ref{app-Whittaker} for more details),
\begin{equation}
\label{A_Whittaker}
f_{\lambda}(z,k)=\frac{1}{\sqrt{2k}}e^{\pi\lambda \xi(t_{\mathrm{h}})/2}W_{\kappa,\mu}(2iz), \qquad \kappa=-i\lambda \xi(t_{\mathrm{h}}), \quad \mu=\frac{1}{2}+s(t_{\mathrm{h}}).
\end{equation}
The derivative of this expression with respect to $z$ can be computed by using Eq.~(\ref{Whittaker-derivative}) in Appendix~\ref{app-Whittaker}.
	
Now we are ready to calculate the boundary terms in Eqs.~(\ref{dot_E_n})--(\ref{dot_B_n}). Let $X$ represent an electromagnetic bilinear function (any of the functions $\mathcal{E}^{(n)}$, $\mathcal{G}^{(n)}$, $\mathcal{B}^{(n)}$), whose spectral decomposition reads	
\begin{equation}
\label{X_integr}
X=\int_{0}^{k_{\mathrm{h}}(t)}\frac{dk}{k}\frac{dX}{d\ln k}.
\end{equation}
Here, $k_{\mathrm{h}}(t)$ is the momentum of the mode that crosses the horizon at time $t$. It can be found by inverting the function $t_{\mathrm{h}}(k)$ in Eq.~(\ref{horizon-crossing-time})
\begin{equation}
\label{k-h}
k_{\mathrm{h}}(t)=\underset{t'\leq t}{\rm max}\Big\{a(t')H(t')\big[|\xi(t')|+\sqrt{\xi^{2}(t')+s^{2}(t')+s(t')}\big]\Big\}.
\end{equation}

The function $X$ varies in time because (i) the integrand evolves in time and (ii) the upper integration limit is time-dependent. The time evolution of the integrand is described by the left-hand side of Eqs.~(\ref{dot_E_n})--(\ref{dot_B_n}). The time dependence of the upper integration limit leads to the boundary terms on the right-hand side of these equations. Thus, the boundary term for $X$ equals its time derivative coming from the variable upper integration limit,
\begin{equation}
\label{X_p_d}
(\dot{X})_{b}= \left. \frac{dX}{d\ln\, k}  \right|_{k=k_{\mathrm{h}}}\cdot\frac{d \ln k_{\mathrm{h}}}{d t}.
\end{equation}
Using the spectral decompositions (\ref{E_1})--(\ref{B_1}) and the approximate solution (\ref{A_Whittaker}) valid in the vicinity of the horizon crossing, we obtain the following boundary terms
\begin{equation}
\label{E_p_d}
[\dot{\mathcal{E}}^{(n)}]_{b}=\frac{d \ln k_{\mathrm{h}}(t)}{d t}\frac{\Delta(t)}{4\pi^{2}}\left(\frac{k_{\mathrm{h}}(t)}{a(t)}\right)^{n+4}\sum_{\lambda=\pm 1}\lambda^{n} E_{\lambda}(\xi(t),s(t)),
\end{equation}
\begin{equation}
\label{G_p_d}
[\dot{\mathcal{G}}^{(n)}]_{b}=\frac{d \ln k_{\mathrm{h}}(t)}{d t}\frac{\Delta(t)}{4\pi^{2}}\left(\frac{k_{\mathrm{h}}(t)}{a(t)}\right)^{n+4}\sum_{\lambda=\pm 1}\lambda^{n+1}G_{\lambda}(\xi(t),s(t)),
\end{equation}
\begin{equation}
\label{B_p_d}
[\dot{\mathcal{B}}^{(n)}]_{b}=\frac{d \ln k_{\mathrm{h}}(t)}{d t}\frac{\Delta(t)}{4\pi^{2}}\left(\frac{k_{\mathrm{h}}(t)}{a(t)}\right)^{n+4}\sum_{\lambda=\pm 1}\lambda^{n}B_{\lambda}(\xi(t),s(t)),
\end{equation}
where $k_{\mathrm{h}}(t)$ is given by Eq.~(\ref{k-h}) and
\begin{equation}
\label{E-lambda}
E_{\lambda}(\xi,s)=\frac{e^{\pi\lambda \xi}}{r^{2}(\xi,s)} \left|\left(i r(\xi,s)-i\lambda \xi-s\right)W_{-i\lambda\xi,\frac{1}{2}+s}(-2i r(\xi,s))+W_{1-i\lambda\xi,\frac{1}{2}+s}(-2i r(\xi,s))\right|^{2},
\end{equation}
\begin{equation}
\label{G-lambda}
G_{\lambda}(\xi,s)=
\frac{e^{\pi\lambda \xi}}{r(\xi,s)} \left\{\Re e\left[W_{i\lambda \xi,\frac{1}{2}+s}(2i r(\xi,s)) W_{1-i\lambda\xi,\frac{1}{2}+s}(-2i r(\xi,s))\right]-s\left|W_{-i\lambda\xi,\frac{1}{2}+s}(-2i r(\xi,s)) \right|^{2}\right\},
\end{equation}
\begin{equation}
\label{B-lambda}
    \qquad B_{\lambda}(\xi,s)=e^{\pi\lambda \xi}\,\left|W_{-i\lambda\xi,\frac{1}{2}+s}(-2i r(\xi,s)) \right|^{2}
\end{equation}
with $r(\xi,s)=|\xi|+\sqrt{\xi^{2}+s+s^{2}}$. For large $\xi$, these expressions can be expanded into series in inverse powers of $r(\xi,s)$, which is convenient for numerical computations. They are listed in Appendix~\ref{app-Whittaker}. We point out again that, in the presence of conductivity (e.g., caused by the Schwinger effect), the evolution of the gauge field becomes nonlocal in time. This nonlocality originates from the exponential factor $\Delta(t)$ in the boundary terms, which contains information about the conductivity in all preceding moments of time, see Eq.~(\ref{Delta-parameter}). This implies that the generation of gauge fields at a given moment of time depends not only on the parameters $H$, $\xi$, and $s$ at the same moment of time, but also on the entire prehistory of the system. We also emphasize once more that the presence of boundary terms is caused by the cutoff in the spectrum of physically relevant gauge-field modes, which changes in time. Moreover, the explicit form of these terms depends on the choice of the cutoff. However, expression (\ref{k-h}) for the cutoff is quite natural, because for any given $k>k_{\mathrm{h}}(t)$, the effective frequency of the mode is always positive for all times $t'\leq t$. Consequently, this mode has not yet experienced any tachyonic instability. This can be clearly seen from the spectra of generated fields. 

\begin{figure}[ht!]
	\centering
	\includegraphics[width=0.97\textwidth]{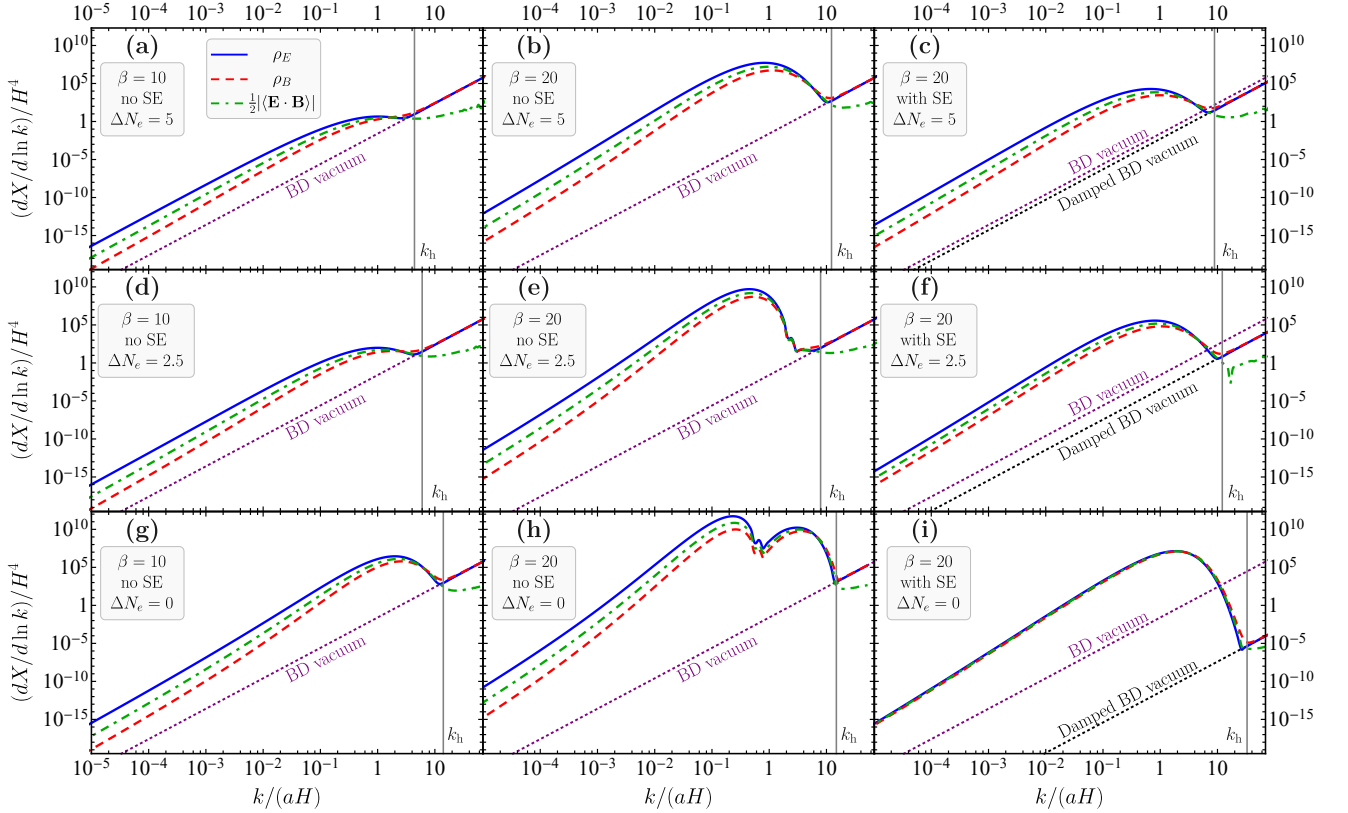}
	\caption{Spectral densities of the electric (blue solid lines) and magnetic (red dashed lines) energy densities as well as of the scalar product $\tfrac{1}{2}|\langle \bm{E}\cdot\bm{B}\rangle|$ (green dashed-dotted lines) at $\Delta N_{e}=5$ (top row), $\Delta N_{e}=2.5$ (middle row), and $\Delta N_{e}=0$ (bottom row) $e$-foldings before the end of inflation.  Gauge fields are generated in the axial-coupling model with the coupling function $I(\phi)=\beta\phi/M_{\mathrm{P}}$ in the three different cases: (left column) $\beta=10$ in the absence of the Schwinger effect; (middle column) $\beta=20$ in the absence of the Schwinger effect; and (right column) $\beta=20$ including the Schwinger effect. Spectra of unperturbed vacuum fluctuations (Bunch-Davies vacuum) are shown by the purple dotted lines. In the presence of the Schwinger effect, subhorizon vacuum fluctuations undergo damping due to the finite conductivity of the medium. The damped Bunch-Davies vacuum spectrum, given by Eq.~(\ref{BD-damped}) is shown by the black dotted lines in panels (c), (f), and (i). The gray vertical lines show the cutoff momenta $k_{\mathrm{h}}$ at the corresponding moments of time.}
	\label{fig-k-h}
\end{figure}

Figure~\ref{fig-k-h} shows the spectral densities of the electric (blue solid lines) and magnetic (red dashed lines) energy densities as well as of the scalar product $\tfrac{1}{2}|\langle \bm{E}\cdot\bm{B}\rangle |$ (green dashed-dotted lines) at $\Delta N_{e}=5$ (top row), 2.5 (middle row), and 0 (bottom row) $e$-foldings before the end of inflation generated in the axial-coupling model $I(\phi)=\beta\phi/M_{\mathrm{P}}$ for different values of the coupling parameter $\beta$ in the absence of the Schwinger effect (left and middle columns) and in its presence (right column). The spectral densities and momenta are normalized by $H^{4}$ and $aH$, respectively, at the corresponding moments of time. The purple dotted lines show the unperturbed spectrum of vacuum fluctuations [corresponding to the Bunch-Davies solution in Eq.~(\ref{BD})]. In the presence of the Schwinger effect, the vacuum fluctuations inside the horizon are damped because of the finite conductivity of the plasma, see Eq.~(\ref{BD-damped}). The corresponding damped Bunch-Davies vacuum spectrum is shown by the black dotted lines in panels (c), (f), and (i). Notice, that the damping is quite strong for large $\beta$, see panel (i). The gray vertical lines show the momenta of the horizon-crossing modes $k_{\mathrm{h}}$ at the corresponding moments of time. This figure nicely illustrates the fact that $k_{\mathrm{h}}$ in form (\ref{k-h}) provides a good separation between the enhanced modes (to the left of the gray vertical lines) and almost unperturbed ones (to the right of the gray vertical lines). Such a separation works well in the absence of backreaction ($\beta=10$, left column) as well as in its presence ($\beta=20$, middle column), and also in the presence of the Schwinger effect (right column). Finally, we would like to comment on the behavior of the spectral density of $\langle \bm{E}\cdot\bm{B}\rangle$, which is shown by the green dashed-dotted lines in Fig.~\ref{fig-k-h}. In the case of a free gauge field, whose mode function is given by the Bunch-Davies solution (\ref{BD}), such a spectral density identically vanishes (while the spectral densities of $\rho_{E}$ and $\rho_{B}$ are equal and scale as $\propto k^{4}$, see the purple dotted lines in Fig.~\ref{fig-k-h}). In the axial-coupling model, we obtain a nonvanishing result for the spectrum of $\langle \bm{E}\cdot\bm{B}\rangle$ for subhorizon modes; however, it is suppressed compared to the spectra of $\rho_{E}$ and $\rho_{B}$. 

\subsection{Truncation of the chain of equations}

As we already mentioned above, the system of equations~(\ref{dot_E_n})--(\ref{dot_B_n}) in general consists of infinitely many equations which form a chain of coupled equations because the equation for the 
quantity of $n$th order contains quantities of order $(n+1)$. However, there is a physical argument which allows us to truncate this chain at some finite order. It can be deduced by considering the spectral decomposition 
(\ref{E_1})--(\ref{B_1}). Any quantity $X^{(n)}$ ($X\in\{\mathcal{E},\, \mathcal{G},\, \mathcal{B}\}$) can be represented in the following form
\begin{equation}
    X^{(n)}=\sum\limits_{\lambda=\pm}\lambda^{n}\int\limits_{0}^{k_{\mathrm{h}}} \Big(\frac{k}{a}\Big)^{n+4} \mathcal{X}_{\lambda}(t,k)\, dk,
\end{equation}
where spectral density function $\mathcal{X}_{\lambda}(t,k)$ does not depend on $n$.
For a sufficiently large $n$, the ultraviolet part of the spectrum in vicinity of the horizon scale $k_{\mathrm{h}}$ gives the dominant contribution to the integral. In this region, we can approximate 
\begin{equation}
    \mathcal{X}_{\lambda}(t,k)\approx \frac{\mathcal{X}_{\lambda}(t)}{a(t)} \Big(\frac{k}{a}\Big)^{p},
\end{equation}
where the factor $1/a$ was added for convenience and $p$ is a spectral index typically of order unity (we take it to be the same for both polarizations because, for the modes which have just crossed the horizon, mode functions behave similarly to the corresponding vacuum solutions which are the same for both polarizations; however, this is not essential for the following discussion). Then, we can approximately write
\begin{equation}
    X^{(n)}\approx \frac{(k_{\mathrm{h}}/a)^{n+p+5}}{n+p+5}\times \sum\limits_{\lambda=\pm}\lambda^{n} \mathcal{X}_{\lambda}(t).
\end{equation}
This expression implies that increasing $n$ by two does not change the second multiplier. Therefore, we have
\begin{equation}
     X^{(n+2)}=\frac{n+p+5}{n+p+7}\Big(\frac{k_{\mathrm{h}}}{a}\Big)^{2} X^{(n)}\approx \Big(\frac{k_{\mathrm{h}}}{a}\Big)^{2} X^{(n)},
\end{equation}
where we assumed $n\gg p=O(1)$. This gives us an opportunity to express higher-order quantities in terms of the lower-order ones and, thus, to truncate the chain. Indeed, if we truncate the system of equations at some $n_{\rm max}$, then the equations of motion for $\mathcal{E}^{(n)}$, $\mathcal{B}^{(n)}$, and $\mathcal{G}^{(n)}$ with $0\leq n \leq (n_{\rm max}-1)$ retain the form of Eqs.~(\ref{dot_E_n})--(\ref{dot_B_n}) and the last three equations for $\mathcal{E}^{(n_{\rm max})}$, $\mathcal{B}^{(n_{\rm max})}$, and $\mathcal{G}^{(n_{\rm max})}$ take the following form
\begin{equation}
\label{dot_E_n_max}
\dot{\mathcal{E}}^{(n_{\rm max})} + [(n_{\rm max}+4)H+2\sigma]\,	\mathcal{E}^{(n_{\rm max})} - 2I'(\phi)\dot{\phi}\,\mathcal{G}^{(n_{\rm max})} +2\Big(\frac{k_{\mathrm{h}}}{a}\Big)^{2}\mathcal{G}^{(n_{\rm max}-1)}=[\dot{\mathcal{E}}^{(n_{\rm max})}]_{b},
\end{equation}
\begin{equation}
\label{dot_G_n_nax}
\dot{\mathcal{G}}^{(n_{\rm max})} +[(n_{\rm max}+4)H+\sigma]\, \mathcal{G}^{(n_{\rm max})}-\Big(\frac{k_{\mathrm{h}}}{a}\Big)^{2}\Big[\mathcal{E}^{(n_{\rm max}-1)}-\mathcal{B}^{(n_{\rm max}-1)}\Big] - I'(\phi)\dot{\phi}\,\mathcal{B}^{(n_{\rm max})}=[\dot{\mathcal{G}}^{(n_{\rm max})}]_{b},
\end{equation}
\begin{equation}
\label{dot_B_n_max}
\dot{\mathcal{B}}^{(n_{\rm max})} + (n_{\rm max}+4)H\,	\mathcal{B}^{(n_{\rm max})}-2\Big(\frac{k_{\mathrm{h}}}{a}\Big)^{2}\mathcal{G}^{(n_{\rm max}-1)}=[\dot{\mathcal{B}}^{(n_{\rm max})}]_{b}.
\end{equation}

\section{Numerical results}
\label{sec-numerical}

The full system of equations describing the generation of gauge fields during axion inflation in the gradient expansion formalism consists of the Friedmann equation (\ref{Friedmann}) for the scale factor, the Klein-Gordon equation (\ref{KGF}) for the inflaton, the system of equations (\ref{dot_E_n})--(\ref{dot_B_n}) for bilinear electromagnetic quantities, which must be truncated at some order $n_{\rm max}$, and the equation for the energy density of charged particles produced by the Schwinger effect (\ref{eq-rho-chi}). In this section, we present our numerical solutions to this system of equations, analyze their accuracy, and compare them to existing results in the literature. 

We consider the generation of the Standard Model hypercharge gauge field axially coupled to the pseudoscalar inflaton field. The axial-coupling function is taken in the simplest linear form
\begin{equation}
\label{axial-coupling-function}
I(\phi)=\frac{\beta}{M_{\mathrm{P}}} \phi
\end{equation}
with one dimensionless coupling parameter $\beta$. Note that the coefficient in front of $\phi$ in the axial coupling function Eq.~(\ref{axial-coupling-function}) is often parametrized as $\alpha/f$, where $f$ is the axion decay constant and $\alpha$ is another dimensionless parameter. However, we consider a generic axionlike inflaton field and do not specify its decay constant $f$; therefore, parametrization (\ref{axial-coupling-function}) is more convenient. 

For numerical analysis, we take the inflaton effective potential in a simple parabolic form
\begin{equation}
\label{inflaton-potential}
    V(\phi)=\frac{M^{2}\phi^{2}}{2}
\end{equation}
with $M=6\times 10^{-6}\,M_{\rm P}$. Although such a potential is already discarded by the CMB observations because it predicts very large magnitude of the tensor-to-scalar power ratio \cite{Planck:2018-infl}, it is still worth considering because many other inflaton potentials can be approximated by Eq.~(\ref{inflaton-potential}) close to their minima. This region appears to be the most important for magnetogenesis. Indeed, the most intensive generation of gauge fields occurs during the last few $e$-foldings of inflation (this is because the generation is determined by the parameter $\xi \propto \dot{\phi}$ and the inflaton velocity $\dot{\phi}$ typically is the largest close to the end of inflation).

The initial condition for the inflaton field $\phi(0)=15.55\,M_{\rm P}$ was chosen to provide at least 60 $e$-foldings of inflation; the initial value of the inflaton velocity was determined from the slow-roll approximation, $\dot{\phi}(0)=-\sqrt{2/3}MM_{\rm P}$. Zero initial values for all electromagnetic quantities $\mathcal{E}^{(n)}$, $\mathcal{B}^{(n)}$, and $\mathcal{G}^{(n)}$ as well as for the energy density $\rho_{\chi}$ of charged fermions were assumed. Moreover, we suppose that the gauge field was also absent before the initial moment of time; consequently, the initial value of the damping parameter $\Delta$ [see Eq.~(\ref{Delta-parameter})] equals unity. We evolve our system of equations from the initial time deep in the inflation stage until its end, where preheating starts and the approximation of a homogeneous inflaton background breaks down. For definiteness, we assume that inflation terminates when the accelerated expansion of the Universe ends, i.e., when the condition $\ddot{a}=0$ is satisfied for the first time.

Numerical values of the parameter $\beta$ typically considered in the literature are of order 10--100. The lower bound is rarely taken to be less than 10 because the gauge field production is very weak in this case. Concerning the upper bound, it is constrained from the requirement that the generated fields do not modify significantly the primordial power spectra at the scales relevant for CMB. The most stringent limit follows from the non-Gaussianity in the scalar power spectrum which results in $|\xi|<2.5$ at the time when the modes relevant for CMB cross the horizon during inflation, i.e., $\delta N_{e}=50-60$ $e$-foldings before the end of inflation (see, e.g.,  Ref.~\cite{Guzzetti:2016}). For the $m^{2}\phi^{2}/2$ inflationary model in the slow-roll approximation, $|\xi|\simeq \beta/\sqrt{2+4 \delta N_{e}}$ that immediately implies $\beta_{\rm max}=|\xi_{\rm max}|\sqrt{2+4 \delta N_{e}}=25-27$ for $\delta N_{e}=50-60$, respectively. Consequently, in our numerical analysis we will restrict ourselves to the values of $\beta$ in the range $10\leq\beta\leq 25$.

\subsection{Small coupling and no Schwinger effect}
\label{subsec-noBR-noSE}

We begin our analysis with the case of small coupling parameter $\beta=10$ in the absence of the Schwinger effect. In such a case, the energy density of the generated gauge field is much less than that of the inflaton; i.e., the gauge field does not backreact on the inflaton evolution and the Universe expansion. This fact allows us to compute the energy densities of the generated gauge fields by an alternative method by using the spectra of generated fields. Indeed, in the absence of the backreaction and Schwinger effect, all Fourier modes of the gauge field evolve independently. Knowing the time dependences of the background inflaton field and the scale factor, we can solve the mode equation (\ref{A_1}) for all physically relevant modes and get the power spectra of the generated fields at any moment of time $t$ [see, e.g., panels (a), (d), and (g) in Fig.~\ref{fig-k-h}]. Then, integrating these spectra over the range of modes which crossed the horizon before the moment of time $t$ during inflation, we get the time dependences of all electromagnetic quantities. These results are obtained without any approximation and, therefore, we use them as reference solutions in order to estimate the accuracy of the results of our gradient expansion formalism.

\begin{figure}[ht!]
	\centering
	\includegraphics[width=0.41\textwidth]{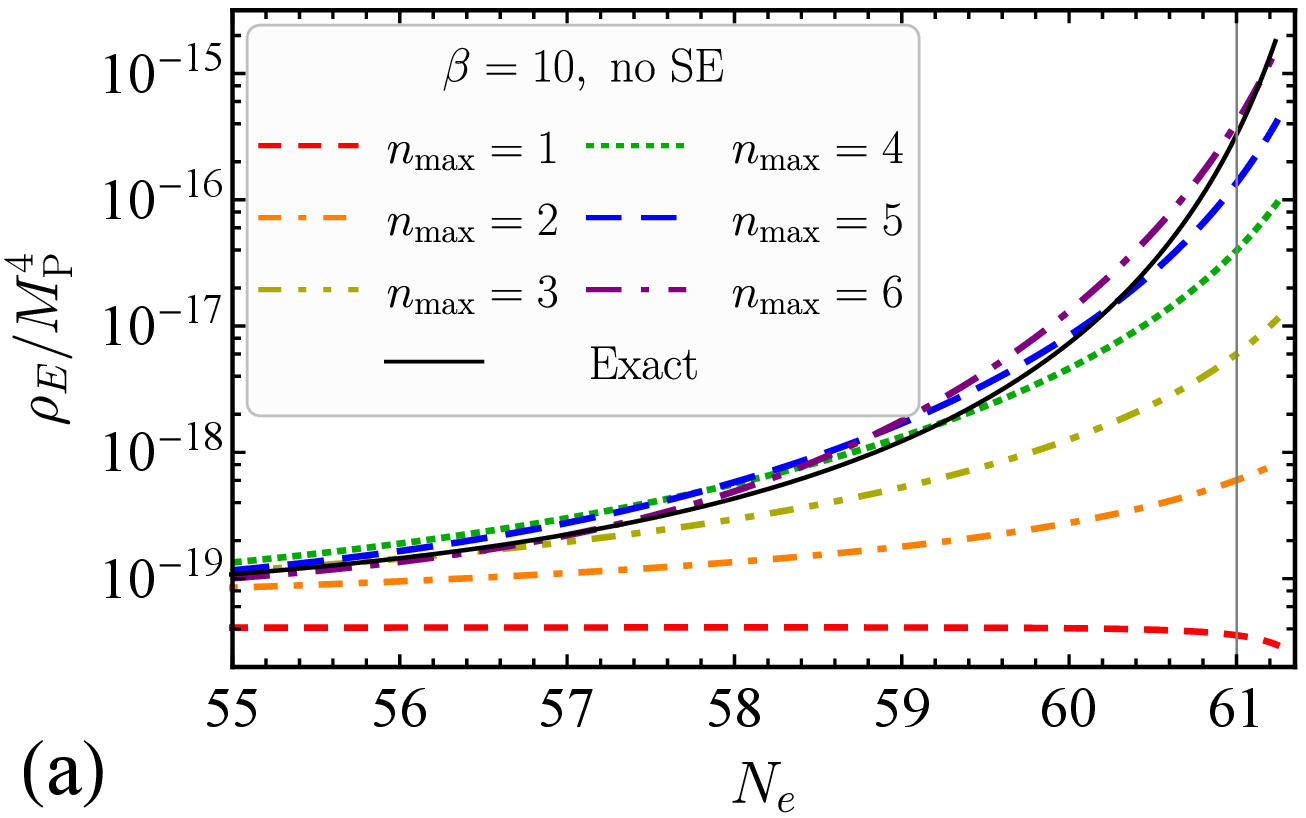}\hspace*{0.5cm}
	\includegraphics[width=0.41\textwidth]{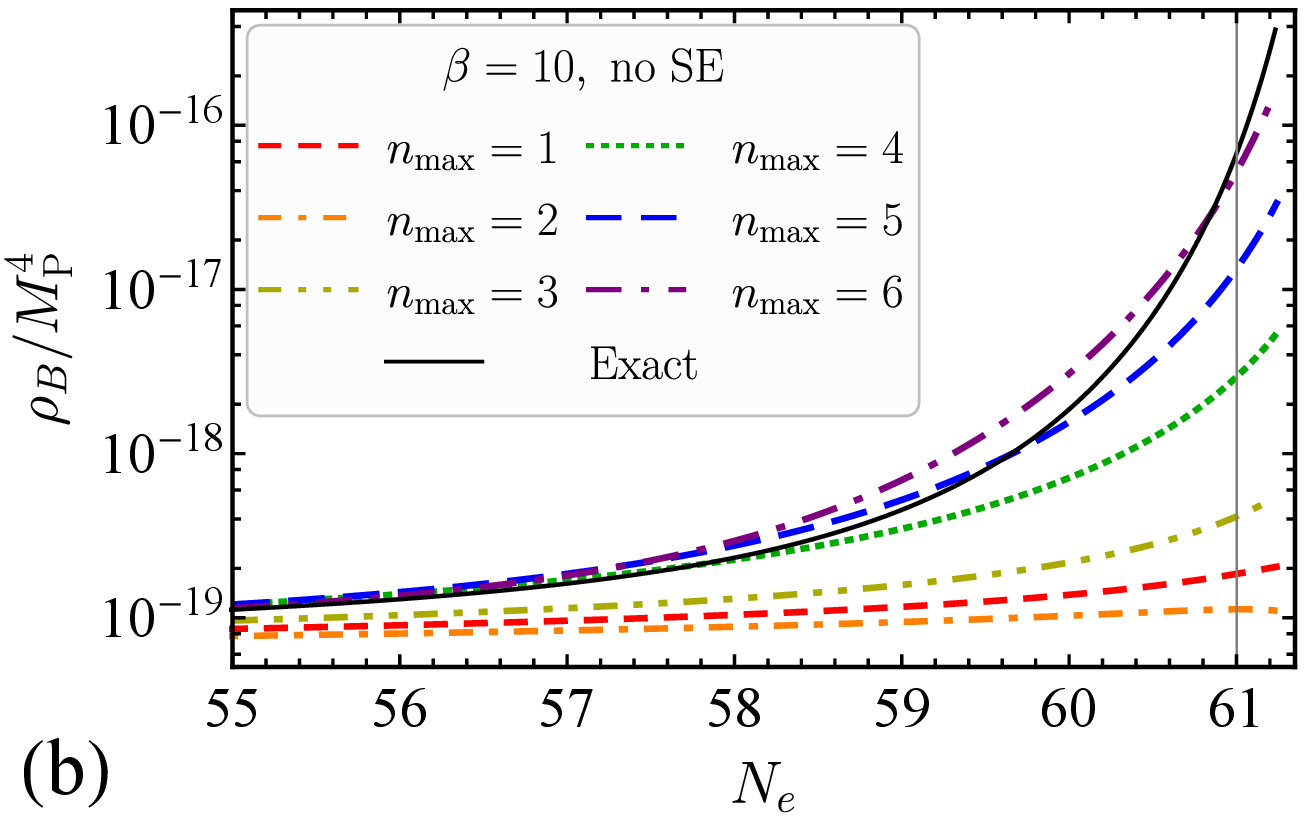}\\
	\includegraphics[width=0.41\textwidth]{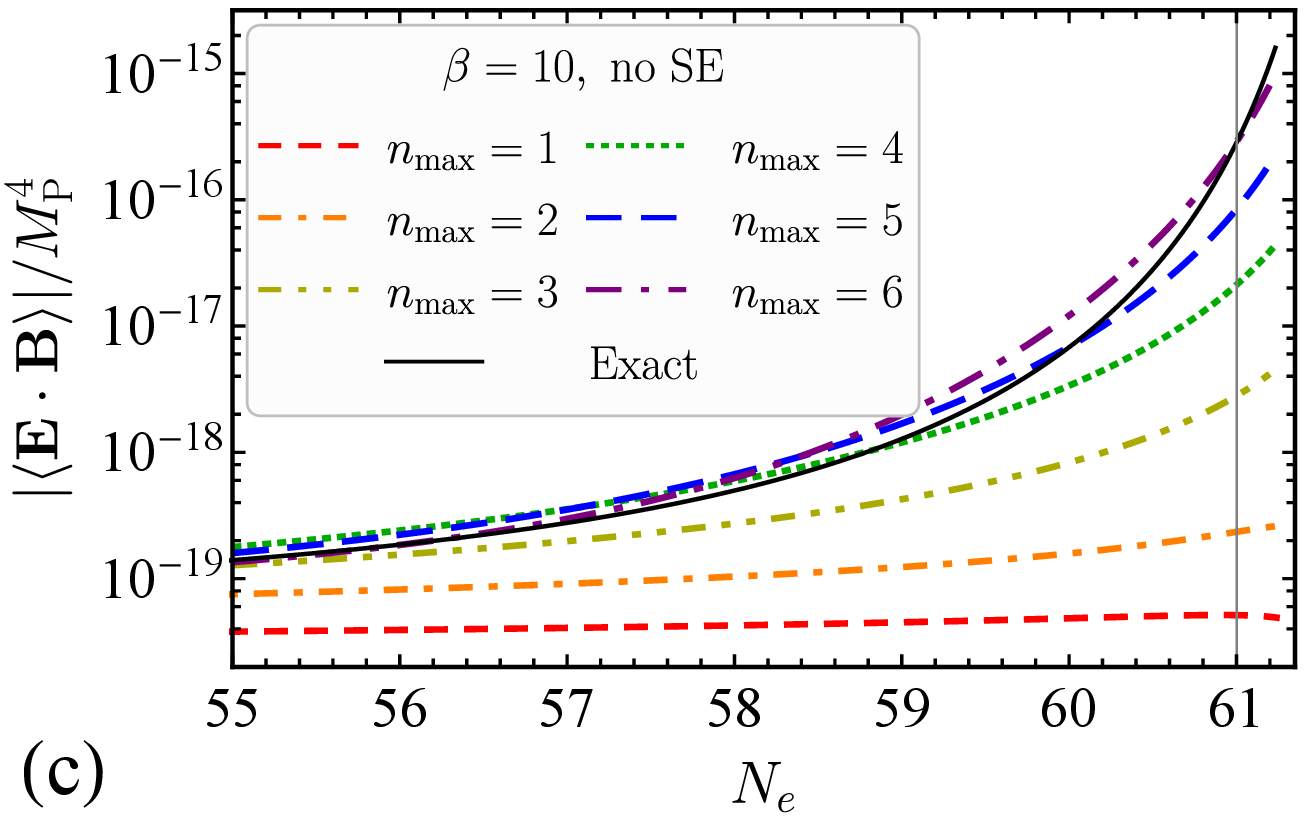}\hspace*{0.8cm}
	\includegraphics[width=0.39\textwidth]{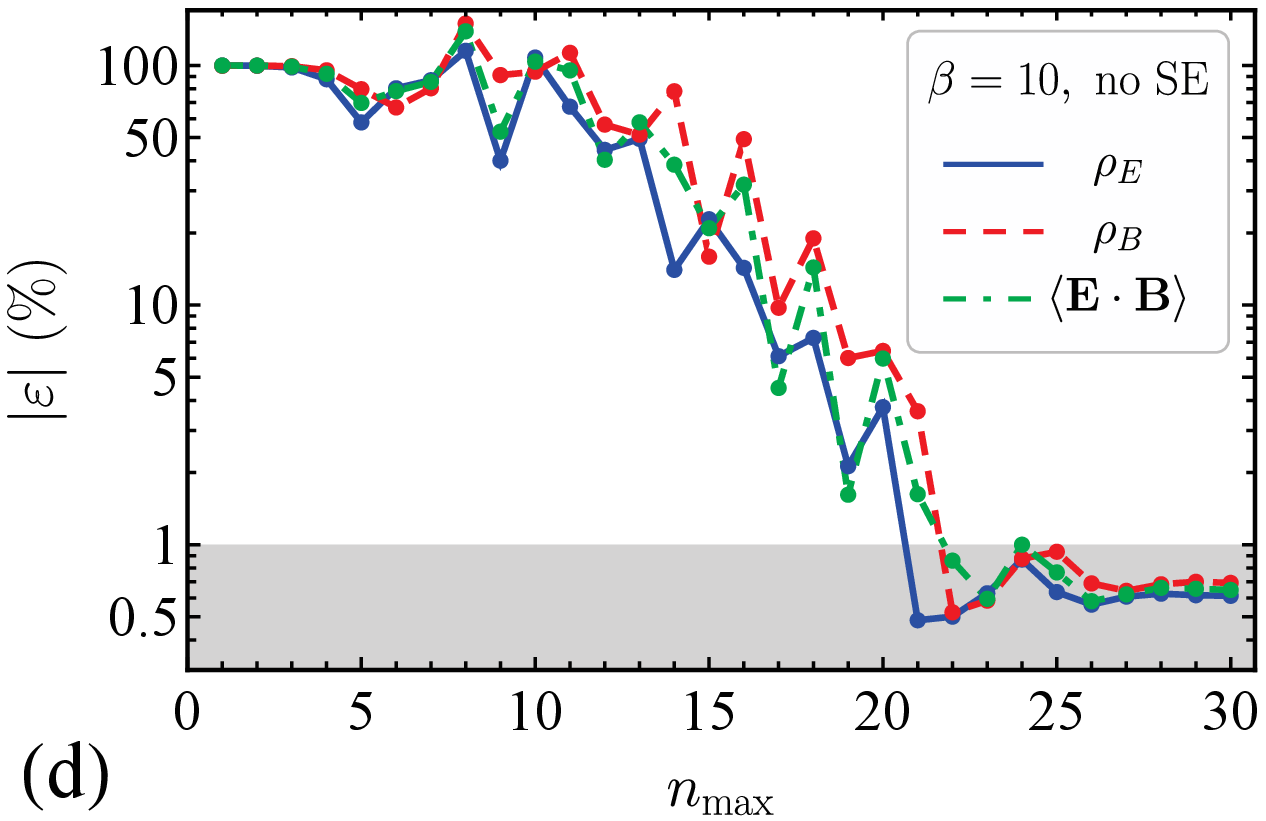}
	\caption{The solutions of the truncated system of equations (\ref{dot_E_n})--(\ref{dot_B_n}) for a few lowest values of the truncation order $n_{\rm max}$ (the colored dashed lines) compared to exact mode-by-mode solutions of Eq.~(\ref{A_1}) shown by the black solid lines. The dependences of (a) the electric energy density $\rho_{E}$, (b) the magnetic energy density $\rho_{B}$, and (c) the scalar product $|\langle\bm{E}\cdot\bm{B}\rangle|$ on the number of $e$-foldings $N_{e}$ from the beginning of inflation in the axial coupling model with $\beta=10$ in the absence of the Schwinger effect. The gray vertical lines mark the end of inflation. (d) The maximal relative error of the approximate results of the gradient expansion formalism for different $n_{\rm max}$ compared to the exact mode-by-mode solution of Eq.~(\ref{A_1}) during inflation. }
	\label{fig-b10-noSE-comparison}
\end{figure}

Figure~\ref{fig-b10-noSE-comparison} illustrates the convergence of the solutions of the system of equations (\ref{dot_E_n})--(\ref{dot_B_n}) truncated at a certain order $n_{\rm max}$ to the exact mode by mode solution with an increase of $n_{\rm max}$. Panels (a), (b), and (c) depict the evolution of the electric, magnetic energy densities, and the scalar product $|\langle\bm{E}\cdot\bm{B}\rangle|$, respectively, during a few last $e$-foldings of inflation (its end is marked by the gray vertical lines on each panel). The dashed curves of different colors show the results for six lowest values of $n_{\rm max}$ while the black solid lines correspond to the reference solution. Panel (d) shows maximal relative errors of the approximate results compared to the mode-by-mode solutions as functions of $n_{\rm max}$. For any electromagnetic quantity $X$, the error is defined in a usual way:
\begin{equation}
\label{error-X}
    \varepsilon_{X}=\frac{X-X_{\rm ref}}{X_{\rm ref}}\times 100\%,
\end{equation}
where $X_{\rm ref}$ is the corresponding reference value. The maximal absolute values of these quantities for $X=\rho_{E}$, $\rho_{B}$, and $\langle\bm{E}\cdot\bm{B}\rangle$ achieved during inflation are shown in Fig.~\ref{fig-b10-noSE-comparison}(d).

\begin{figure}[ht!]
	\centering
	\includegraphics[width=0.42\textwidth]{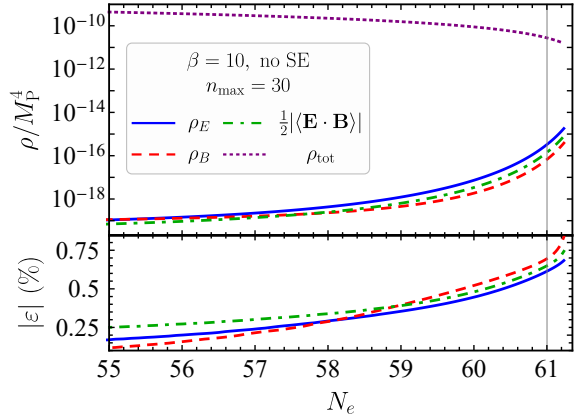}
	\caption{The evolution of the electric (blue solid line) and magnetic (red dashed line) energy densities as well as the scalar product $\tfrac{1}{2}|\langle\bm{E}\cdot\bm{B}\rangle|$ (green dashed-dotted lines) during the last few $e$-foldings of inflation in the axial coupling model with $\beta=10$ in the absence of the Schwinger effect. The top plot presents the numerical results obtained form the system of equations (\ref{dot_E_n})--(\ref{dot_B_n}) truncated at $n_{\rm max}=30$ while the bottom plot shows the relative error of this solution compared to the exact mode-by-mode solution of Eq.~(\ref{A_1}). The purple dotted line shows the total energy density of the Universe. The gray vertical line marks the end of the inflation stage.}
	\label{fig-b10-noSE-rho}
\end{figure}

First of all, we should mention that there is a general tendency for the error to decrease with increasing $n_{\rm max}$. However, this decrease is nonmonotonic; for some specific values of $n_{\rm max}$ (e.g., 9, 17, 19) the error is less than for both neighboring orders; for some other values (e.g., 8, 18, 20) the situation is opposite. Finally, for $n_{\rm max}\gtrsim 25$ the error ceases to decrease and stabilizes at the level of $0.6\%-0.8\%$. This irremovable error is not caused by the truncation procedure as it does not depend on the truncation order. It comes from other approximations used in the derivation of the boundary terms; e.g., from assumptions of constant $\xi$ and exact de Sitter universe expansion in Eq.~(\ref{A_3}). This can be confirmed by the fact that the residual error increases during the last $e$-foldings of inflation (see, the bottom plot in Fig.~\ref{fig-b10-noSE-rho}) when the above mentioned approximations become less justified. Nevertheless, the achieved accuracy is absolutely satisfactory for all physical applications of the results.

The energy densities of the generated gauge field during the last few $e$-foldings of inflation compared to the total energy density of the Universe are shown in the top plot of Fig.~\ref{fig-b10-noSE-rho}. They were calculated in the gradient expansion formalism with $n_{\rm max}=30$, and are in good accordance with the exact results during the whole inflation stage (see the bottom plot of Fig.~\ref{fig-b10-noSE-rho} for the corresponding relative error of the result). This figure confirms that the generated fields are weak (the corresponding energy density is $4-5$ orders of magnitude less than that of the inflaton) and cannot backreact on the Universe expansion. 
Close to the end of inflation, the electric and magnetic fields become almost collinear. This follows from the relative position of the curves for $\rho_{E}$ (blue solid line), $\rho_{B}$ (red dashed line), and $\tfrac{1}{2}|\langle\bm{E}\cdot\bm{B}\rangle|$ (green dashed-dotted line). In the logarithmic scale, the third one is located almost in the middle between the first two; i.e., $|\langle\bm{E}\cdot\bm{B}\rangle|\approx \sqrt{\langle\bm{E}^{2}\rangle \langle\bm{B}^{2}\rangle}$ (more precisely, $\cos\theta$ grows from 0.86 to 0.96 during the last two $e$-foldings of inflation). This is a consequence of the fact that in the process of enhancement the gauge field becomes more helical and estimate (\ref{estimate-theta}) becomes valid.

\subsection{Large coupling and no Schwinger effect}
\label{subsec-BR-noSE}

Let us now consider larger values of the coupling parameter, $\beta=20$ and 25, still in the absence of the Schwinger effect. In this case, the generated gauge field backreacts on the background evolution. First of all, the term on the right-hand side of Eq.~(\ref{KGF}) becomes important and slows down the inflaton rolling. Second, the energy density of the gauge field becomes comparable to that of the inflaton and impacts the expansion rate of the Universe. Consequently, we cannot apply the standard spectral approach and find the solution of the mode equation (\ref{A_1}) for each separate mode on a given inflaton background because this background itself depends on the resulting gauge field. Thus, all modes of the gauge field become coupled due to backreaction. One possible solution to this problem is to apply the iterative procedure which converges to a self-consistent configuration of the gauge field and the inflaton \cite{Domcke:2020}. 

Another possibility is to use the gradient expansion formalism since it takes into account all relevant modes of the gauge field at once. As already emphasized, its main advantage is that it does not require any iterative procedure and allows us to get the results after a single numerical run. First, we find numerically the truncation order $n_{\rm max}$ starting from which the solution of the truncated system of equations (\ref{dot_E_n})--(\ref{dot_B_n}) does not change with increasing $n_{\rm max}$. Thus, we minimize the error caused by the truncation procedure. As a result, we obtain time dependences of all electromagnetic bilinear functions as well as the inflaton and scale factor. Second, we use the last two quantities to solve the mode equation (\ref{A_1}) and get the gauge field power spectra [for $\beta=20$ see, e.g., panels (b), (e), and (h) in Fig.~\ref{fig-k-h}]. Finally, integrating these spectra over the momenta of all relevant modes, we compute the electromagnetic bilinear functions. These solutions cannot be considered as the exact ones because they are based on the background values of the inflaton and scale factor which have been found by an approximate method.
Nevertheless, we can use them as reference solutions in order to check the consistency of the gradient expansion formalism. Quantitatively it can be characterized by the relative deviation between the two solutions given by Eq.~(\ref{error-X}), where $X$ is the solution found from the gradient-expansion approach and $X_{\rm ref}$ is the corresponding result calculated from the gauge-field spectrum.

\begin{figure}[ht!]
	\centering
	\includegraphics[width=0.42\textwidth]{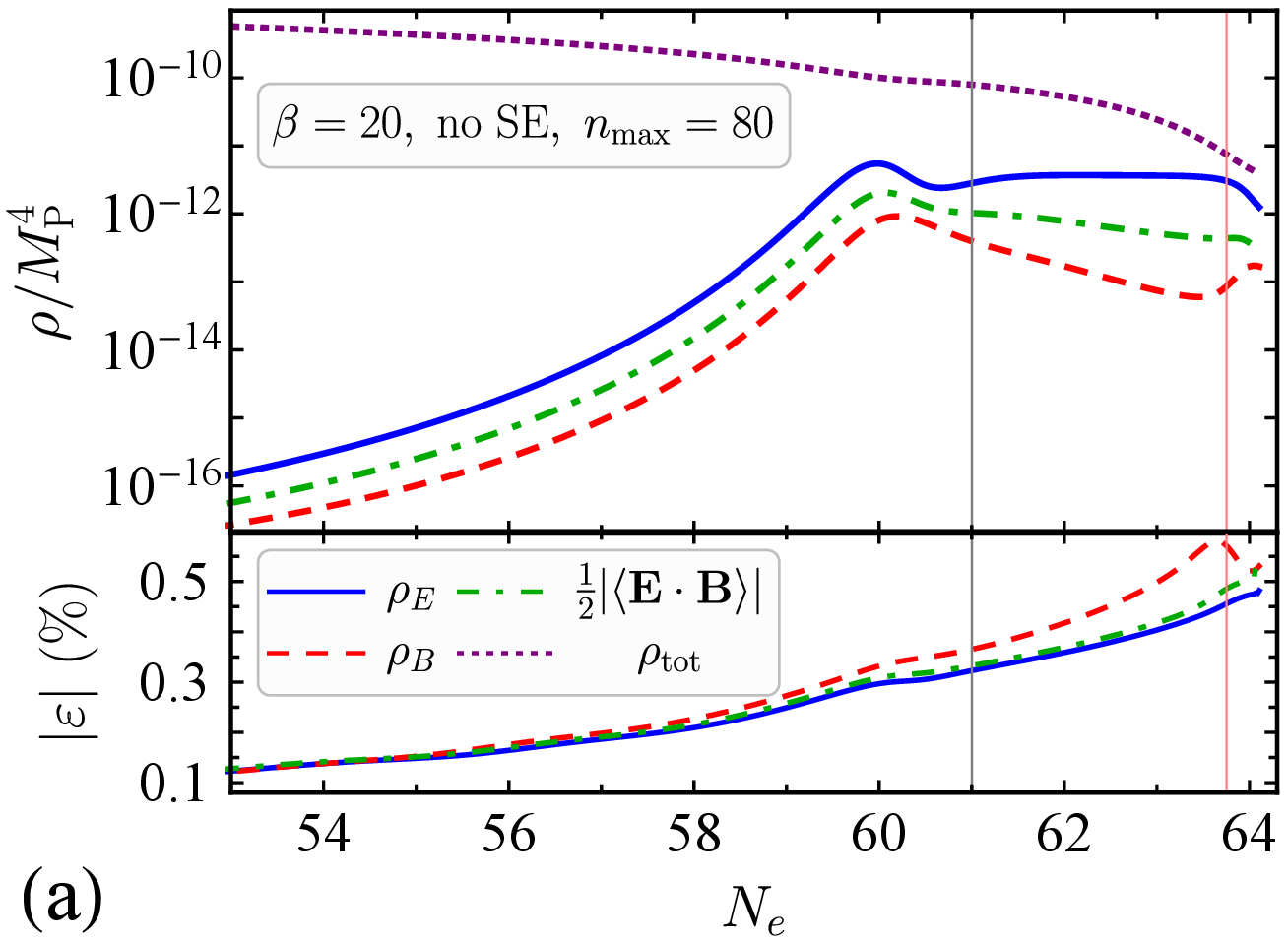}\hspace*{0.7cm}
	\includegraphics[width=0.42\textwidth]{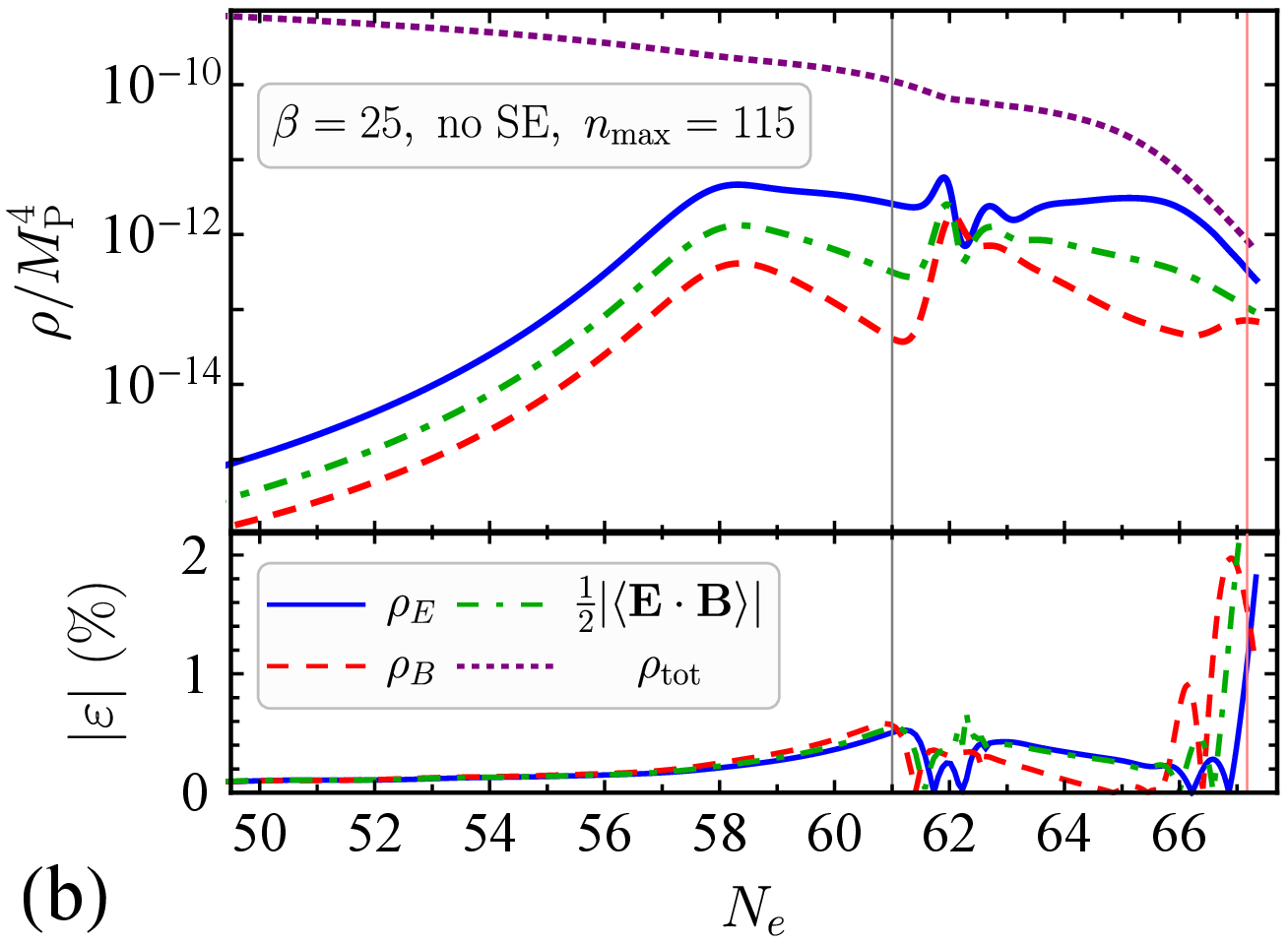}
	\caption{The dependences of the electric energy density (blue solid lines), magnetic energy density (red dashed lines), and the scalar product $\tfrac{1}{2}|\langle\bm{E}\cdot\bm{B}\rangle|$ (green dashed-dotted lines) on the number of $e$-foldings close to the end of inflation in the axial coupling model with (a) $\beta=20$ and (b) $\beta=25$ in the absence of the Schwinger effect. The purple dotted lines show the total energy density of the Universe. The energy density of the generated gauge field is large enough to cause the backreaction which slows the rolling of the inflaton and extends the inflation stage. The gray vertical lines show the end of inflation in the absence of backreaction while the pink vertical lines mark the actual end of inflation in each case. The top plots show the numerical results obtained from the system of equations (\ref{dot_E_n})--(\ref{dot_B_n}) truncated at (a) $n_{\rm max}=80$ and (b) $n_{\rm max}=115$ while the bottom plots show the relative deviation of this solution from the mode by mode solution of Eq.~(\ref{A_1}) on the inflaton background modified by the backreaction.}
	\label{fig-b20-25-noSE}
\end{figure}

Figure~\ref{fig-b20-25-noSE} demonstrates the evolution of the electric (blue solid lines) and magnetic (red dashed lines) energy densities as well as the scalar product $\tfrac{1}{2}|\langle\bm{E}\cdot\bm{B}\rangle|$ (green dashed-dotted lines) during the last few $e$-foldings of inflation for $\beta=20$ [panel (a)] and $\beta=25$ [panel (b)]. The total energy density of the Universe is shown by the purple dotted lines. The bottom plots show the relative deviation of the obtained solutions from the reference ones. For $\beta=20$, this deviation is always less than 0.6\%. This is also true for $\beta=25$ until the last $e$-folding of inflation where the deviation increases to $\sim 2\%$. For the majority of physical applications such an accuracy of the solution should be absolutely sufficient. However, it can be further increased applying the iterative procedure of Ref.~\cite{Domcke:2020}.

\begin{figure}[ht!]
	\centering
	\includegraphics[width=0.42\textwidth]{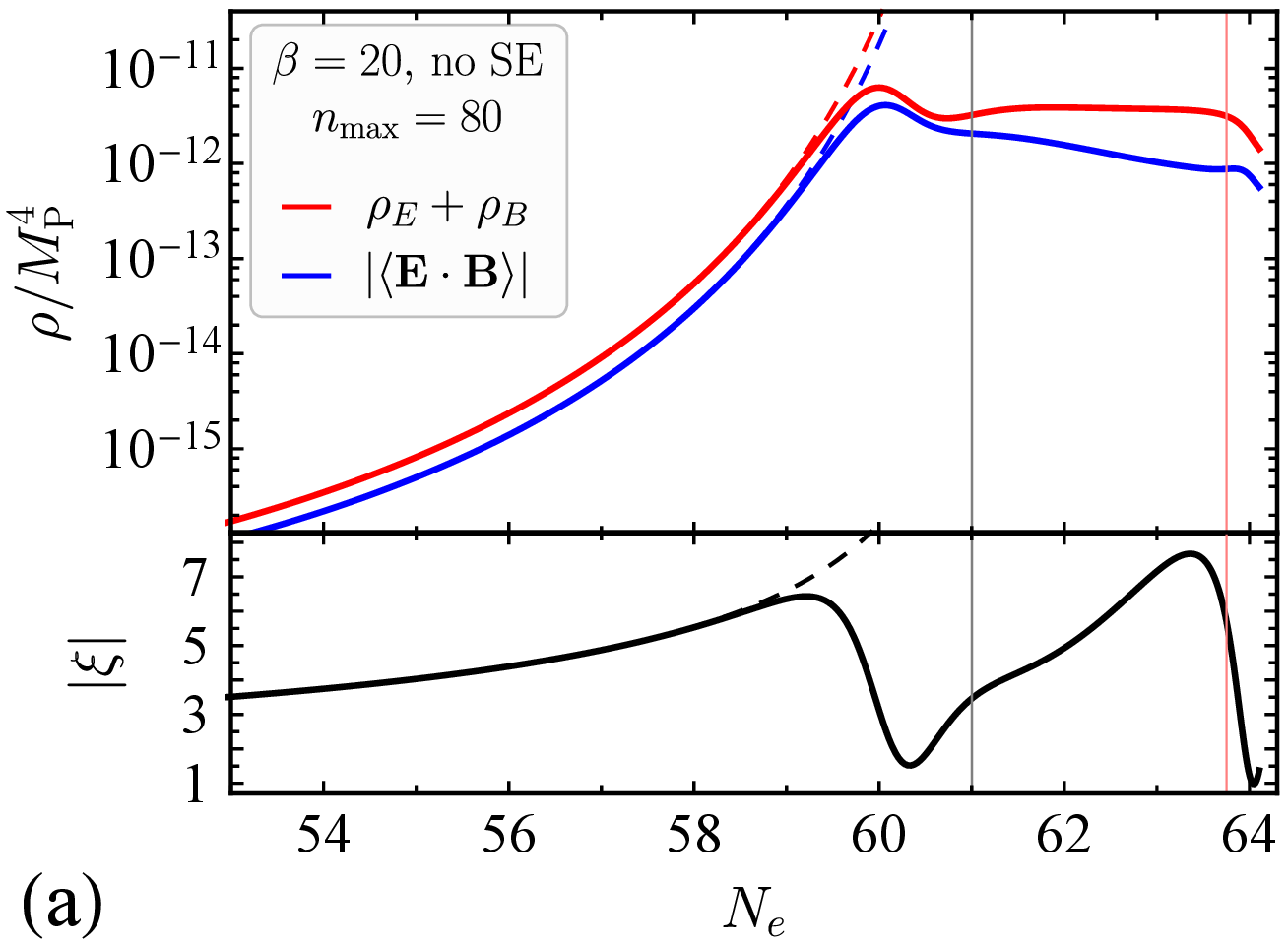}\hspace*{0.7cm}
	\includegraphics[width=0.42\textwidth]{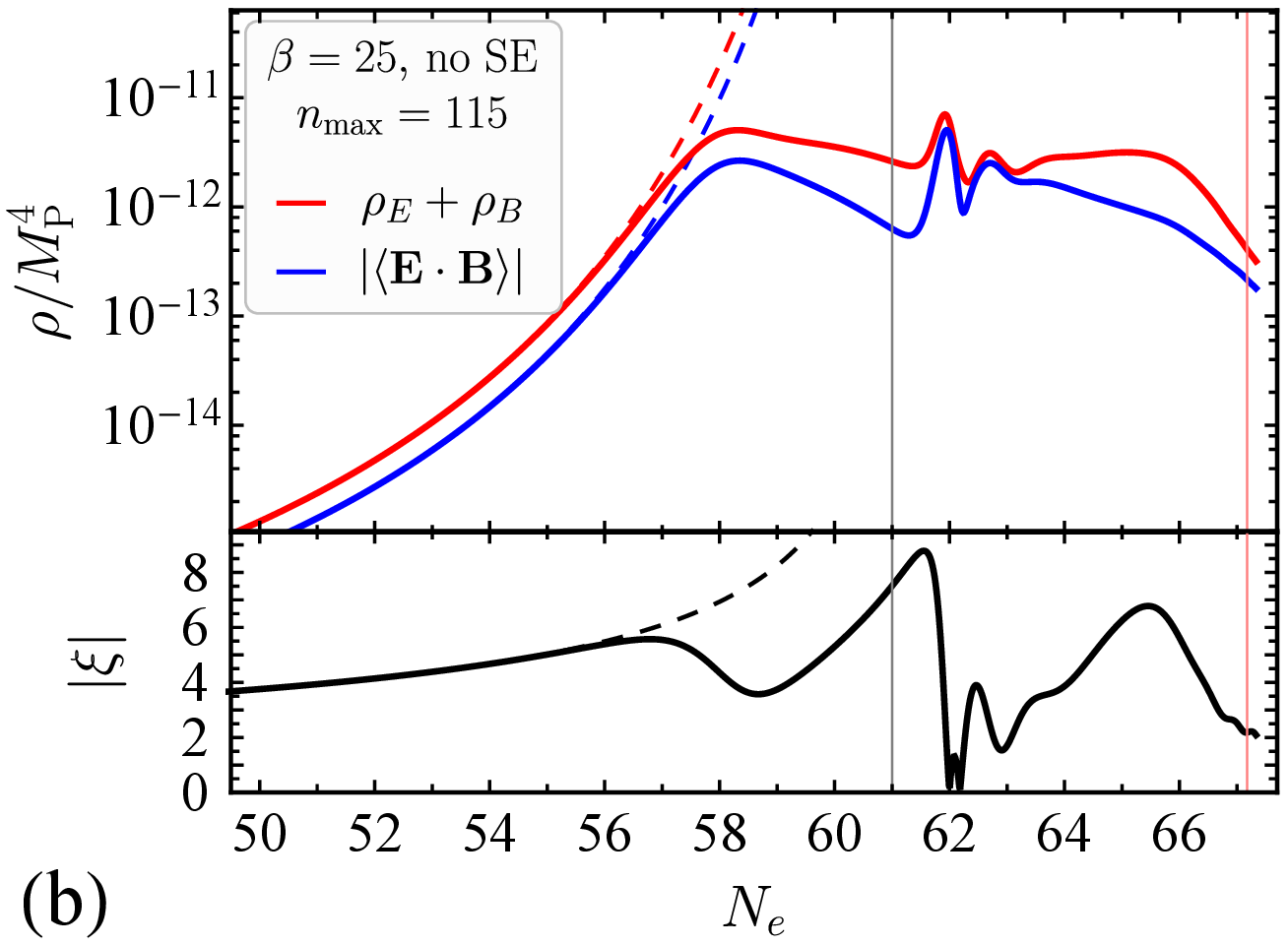}
	\caption{Top plots: the total electromagnetic energy density (red solid lines) and the scalar product $|\langle\bm{E}\cdot\bm{B}\rangle|$ (blue solid lines) as functions of $N_{e}$ generated in the axial coupling model with (a) $\beta=20$ and (b) $\beta=25$ in the absence of the Schwinger effect. Bottom plots show the absolute value of the parameter $\xi$. The corresponding dashed lines show the same dependences in the absence of backreaction. The pink vertical lines mark the end of inflation in each case while the gray vertical lines show the end of inflation in the absence of backreaction. These solutions obtained from the gradient expansion formalism are in good accordance with the results of the iterative solution of the mode equation (\ref{A_1}), presented in Ref.~\cite{Domcke:2020}, cf. Fig.~6 there.}
	\label{fig-b20-25-noSE-rho-xi}
\end{figure}

In order to analyze the evolution of the system in more detail, we show the total electromagnetic energy density (red line), the scalar product $|\langle\bm{E}\cdot\bm{B}\rangle|$ (blue line) and the parameter $\xi$, which is responsible for the gauge field generation, in Fig.~\ref{fig-b20-25-noSE-rho-xi}. The dashed lines of the same colors represent the corresponding quantities in the absence of backreaction (when the inflaton background remains unperturbed). At first, the gauge field energy density monotonically increases in time because the parameter $\xi$ constantly grows during inflation in the absence of the backreaction. However, when the impact of generated fields on the inflaton dynamics becomes important, the growth of the energy densities abruptly stops and the gauge field evolution enters the nonlinear stage. In this strong backreaction regime, the energy densities, the scalar product $\langle\bm{E}\cdot\bm{B}\rangle$, as well as the parameter $\xi$ exhibit oscillatory features previously reported in the literature \cite{Cheng:2015,Notari:2016,Domcke:2020} (for easier comparison with the previously reported results, we plotted Fig.~\ref{fig-b20-25-noSE-rho-xi} for the same values of the parameters and in the same style as Fig.~6 in Ref.~\cite{Domcke:2020}).
Moreover, since the inflaton rolls much slower, the inflation stage becomes a few $e$-foldings longer, which is clearly seen from Figs.~\ref{fig-b20-25-noSE} and \ref{fig-b20-25-noSE-rho-xi}, where the actual end of inflation [determined from the condition $\ddot{a}(t)=0$] is shown by the pink vertical lines and the end of inflation without the backreaction is marked by the gray vertical lines. Our gradient expansion formalism allows to recover all features of the backreaction regime previously reported in the literature.

\subsection{Small and large coupling with the Schwinger effect included}
\label{subsec-SE}

Finally, let us consider the full physical system which includes also charged fermions produced due to the Schwinger effect during axion inflation. In such a case, even in the absence of backreaction the gauge field evolves in the nonlinear regime, because the Schwinger conductivity entering the mode equation~(\ref{A_1}) depends on the total electromagnetic field; i.e., it couples all relevant modes to each other. This does not allow us to determine the exact solution for the generated gauge field by solving the mode equation (\ref{A_1}) separately for each mode as we did in Sec.~\ref{subsec-noBR-noSE}. This can be done, e.g., by using the iterative approach which still has not been realized in the literature. Instead, we perform the same procedure as discussed in Sec.~\ref{subsec-BR-noSE}; i.e., we take the stable solution obtained from the gradient expansion approach and use it to solve the mode equation~(\ref{A_1}). Electromagnetic bilinear quantities then can be found from the spectra and compared to the results of the gradient expansion formalism. The relative deviation between them, given by Eq.~(\ref{error-X}), characterizes the consistency of our approach.

\begin{figure}[ht!]
	\centering
	\includegraphics[height=0.31\textwidth]{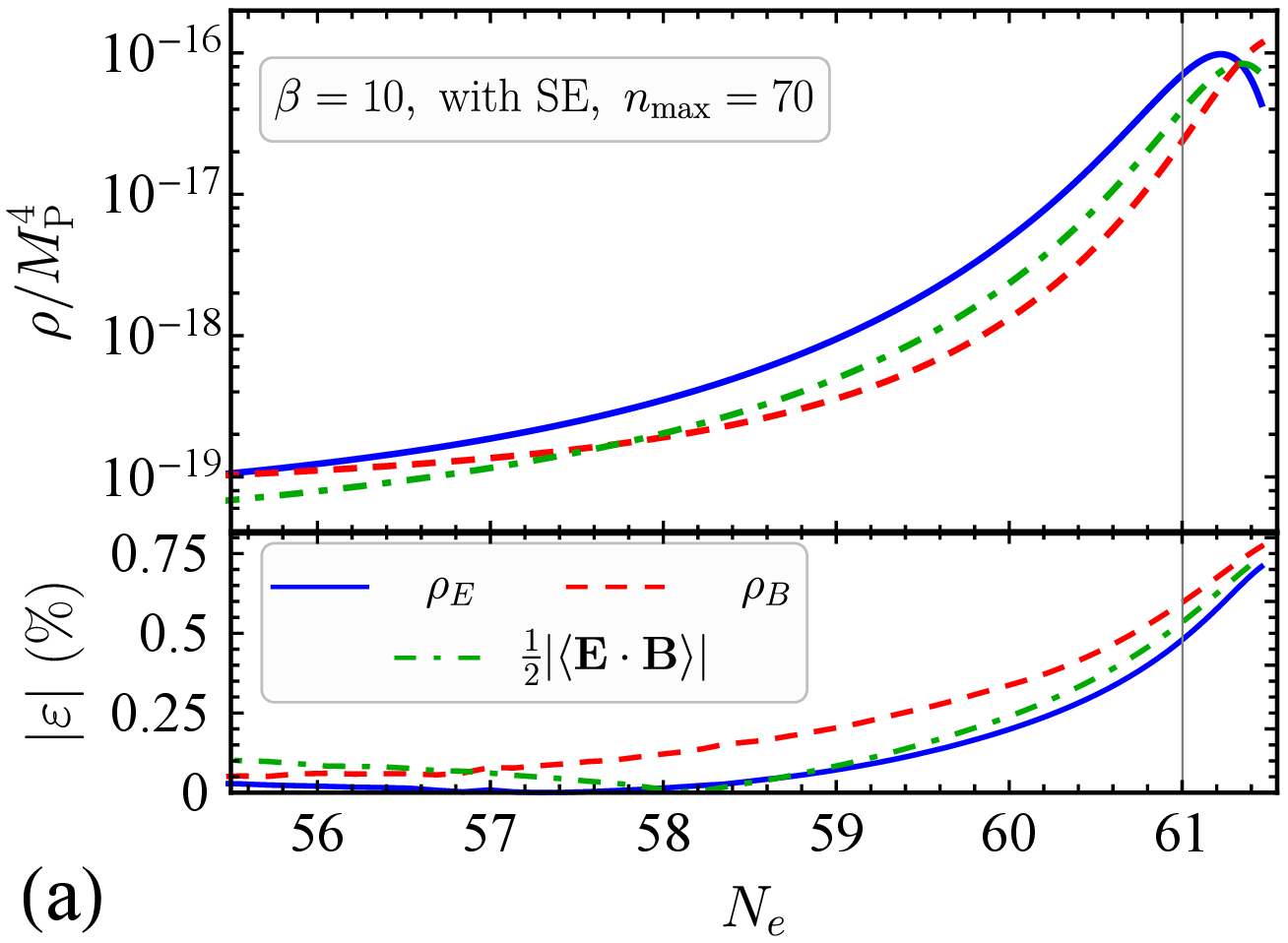}\hspace*{0.7cm}
	\includegraphics[height=0.31\textwidth]{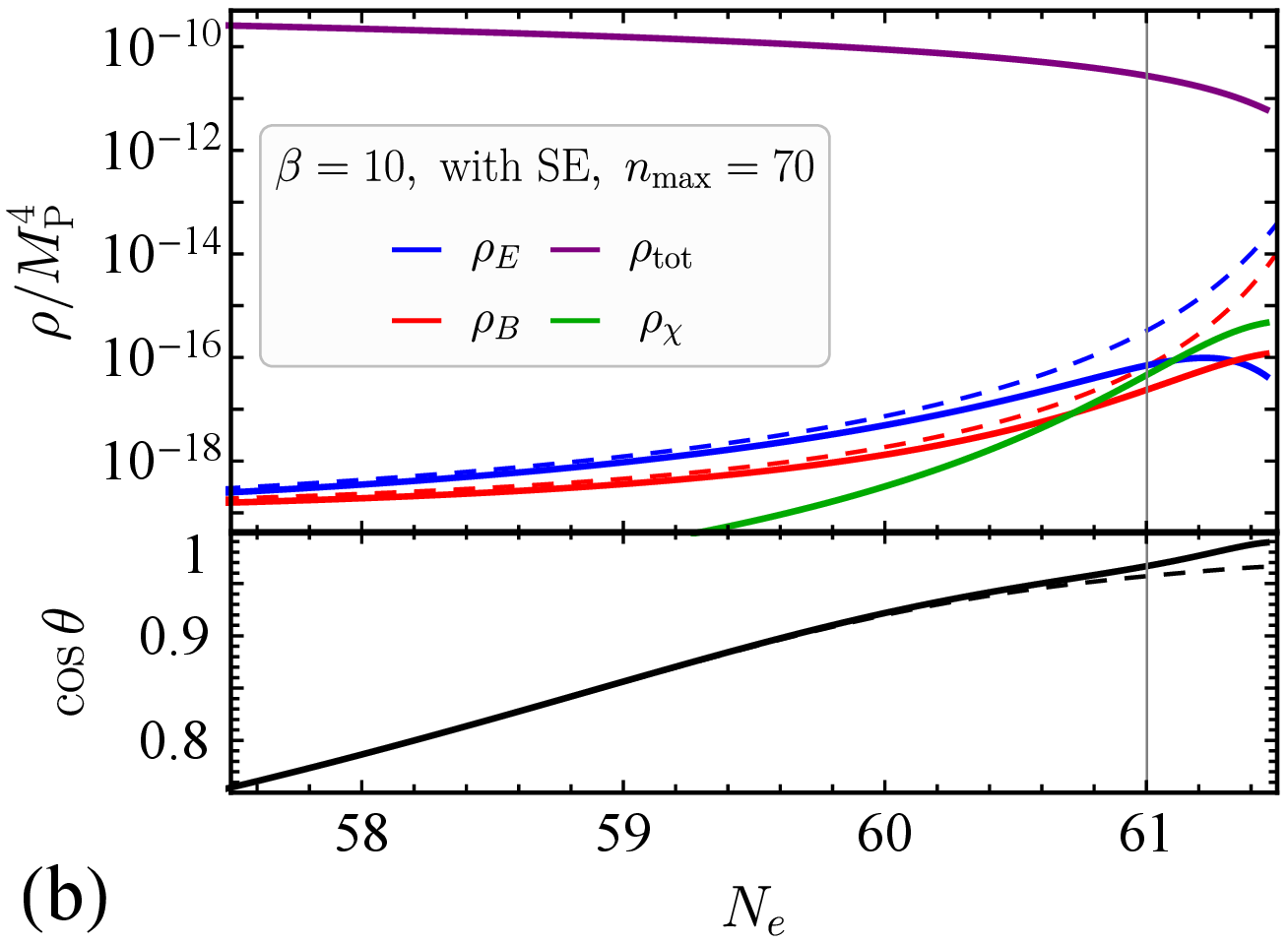}
	\caption{(a) The electric energy density (blue solid line), magnetic energy density (red dashed line) and the scalar product $\tfrac{1}{2}|\langle\bm{E}\cdot\bm{B}\rangle|$ (green dashed-dotted line) generated in the axial coupling model with $\beta=10$ taking into account the Schwinger production of all species of the massless Standard Model fermions. The top plot shows the results obtained from the system of equations (\ref{dot_E_n})--(\ref{dot_B_n}) truncated at $n_{\rm max}=70$ while the bottom plot shows their relative error compared to the results of mode-by-mode solution of Eq.~(\ref{A_1}) on the background of the inflaton and the Schwinger conductivity determined from the gradient expansion formalism. (b) The components of the energy density of the Universe during the last few $e$-foldings of inflation: the electric energy density (blue solid line), magnetic energy density (red solid line), energy density of charged fermions produced by the Schwinger effect (green solid line), and the total energy density of the Universe (purple solid line). The bottom plot shows the cosine of the angle $\theta$ between the electric and magnetic fields. The dashed lines show the corresponding dependences in the absence of the Schwinger effect. The gray vertical lines in both panels mark the end of inflation.}
	\label{fig-b10-SE}
\end{figure}

\begin{figure}[ht!]
	\centering
	\includegraphics[height=0.31\textwidth]{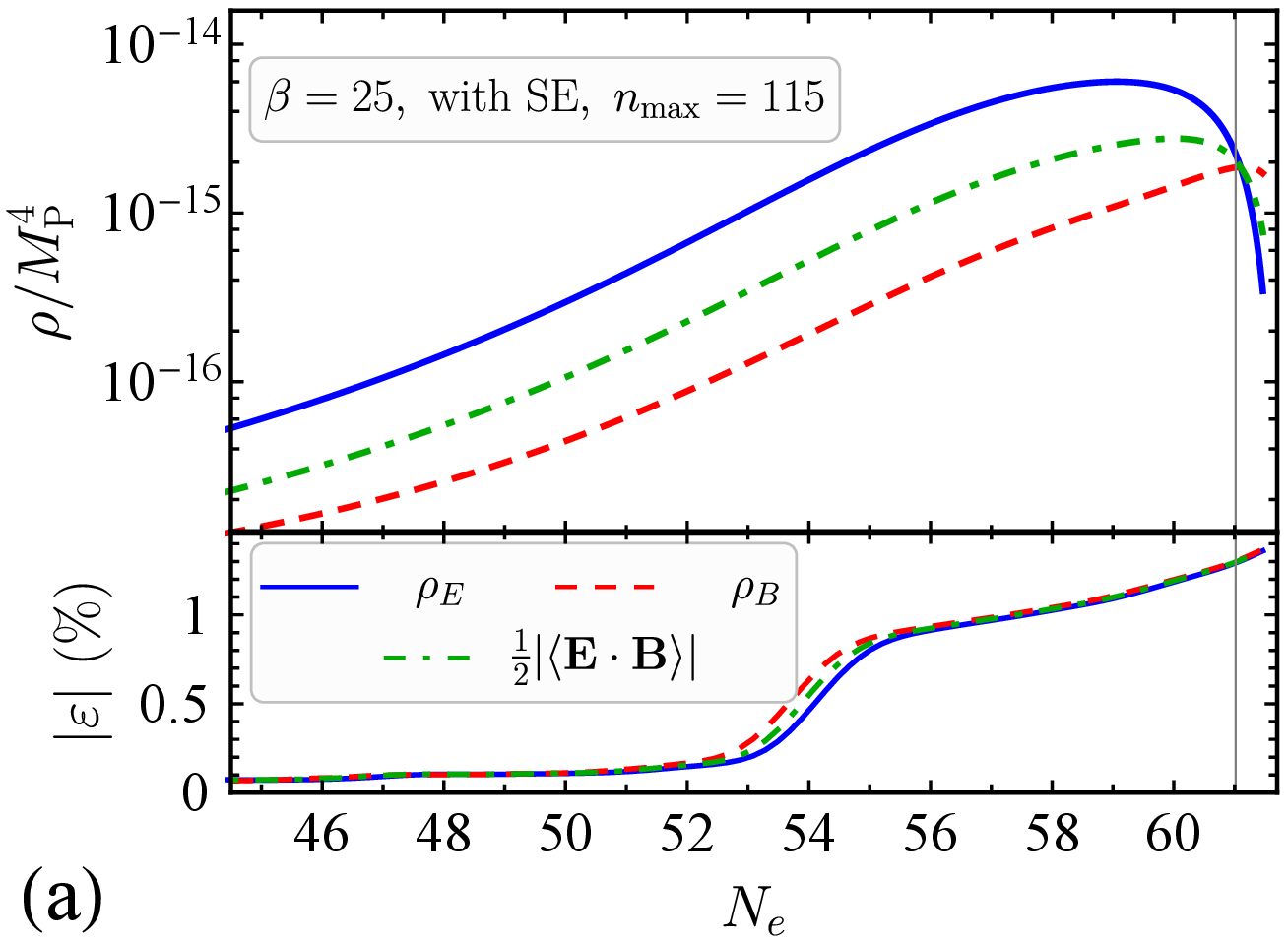}\hspace*{0.7cm}
	\includegraphics[height=0.31\textwidth]{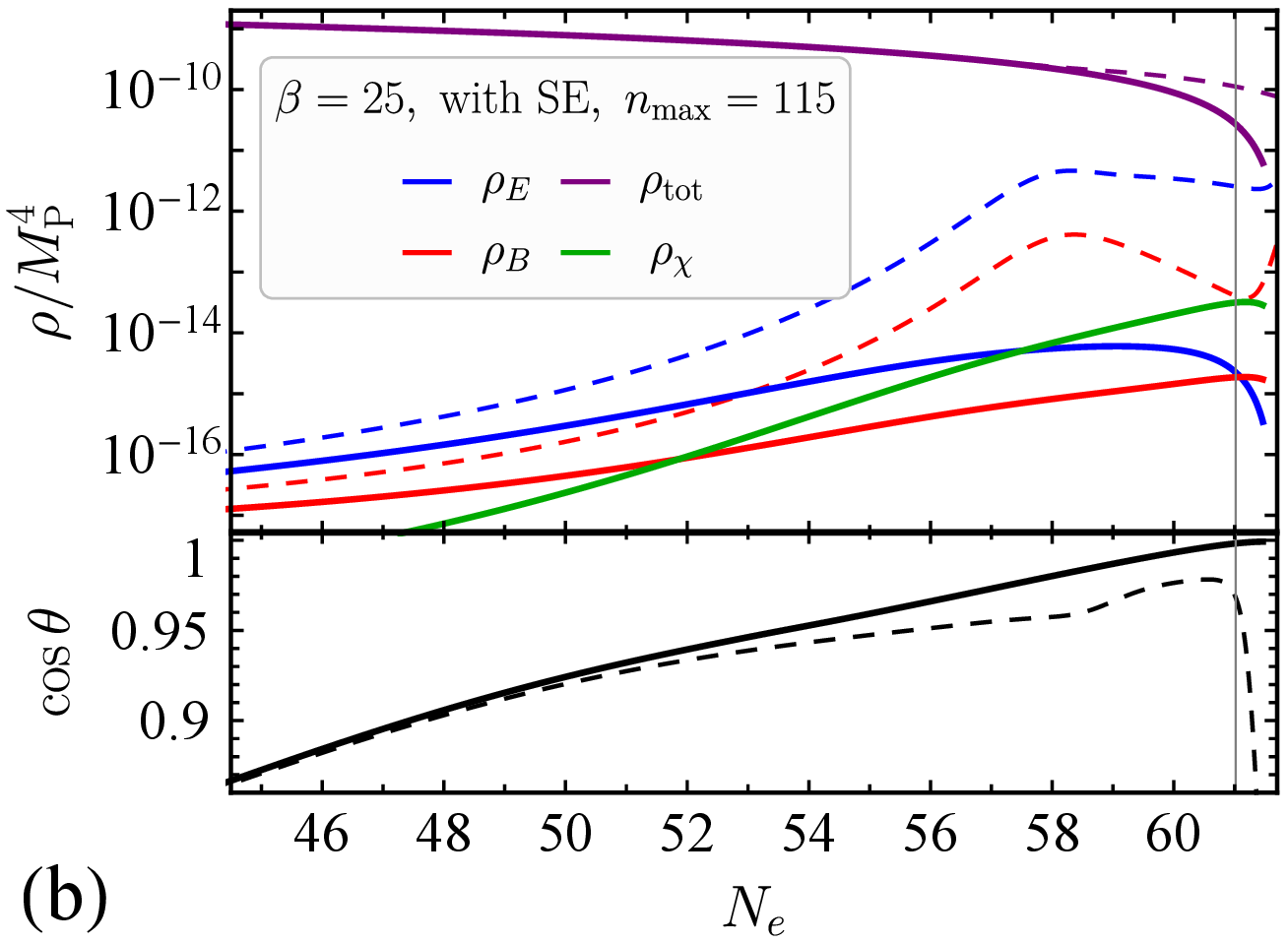}
	\caption{The same quantities as shown in Fig.~\ref{fig-b10-SE}(a) and \ref{fig-b10-SE}(b), respectively, for the case of $\beta=25$, $n_{\rm max}=115$.}
	\label{fig-b25-SE}
\end{figure}

We present the results of our gradient expansion approach in Figs.~\ref{fig-b10-SE} and \ref{fig-b25-SE} for $\beta=10$ and 25, respectively. Panels (a) show the dependences of the electric (blue solid lines), magnetic (red dashed lines) energy densities, and the scalar product $\tfrac{1}{2}|\langle\bm{E}\cdot\bm{B}\rangle|$ (green dashed-dotted line) on the number of $e$-foldings close to the end of axion inflation. Bottom plots show relative deviations from the reference solutions found from the spectra. As in the previous subsections, these deviations are of order 1\% meaning that our gradient expansion formalism allows to get a highly self-consistent result even in the most complicated case when both backreaction and Schwinger effect are taken into account. Moreover, such an accuracy is achieved in a single run of the code without iterative procedure. 

The top plots in Figs.~\ref{fig-b10-SE}(b) and \ref{fig-b25-SE}(b) represent different components of the energy density of the Universe: electric (blue lines), magnetic (red lines), charged particles (green lines), and the total energy density (purple lines). Dashed lines of the respective color show similar dependences in the absence of the Schwinger effect (considered in more detail in previous subsections). Comparing the solid and dashed lines it is easy to conclude that the Schwinger effect significantly suppresses magnetogenesis. In the case of large coupling parameter $\beta=25$, the gauge-field energy density becomes three orders of magnitude less than in the absence of the Schwinger effect and does not cause the backreaction any more. As a result, the inflation stage has the same duration as in the unperturbed case. Therefore, for typical values of the coupling parameter $\beta$ considered in the literature, the backreaction does not occur in the presence of the Schwinger effect. Unless we take extremely large values of $\beta$ or larger values of $H$, we do not expect any backreaction.

Especially strong suppression occurs for the electric component of the energy density, because it is directly affected by the conductivity, see Eq.~(\ref{dot_E_n}). Although it dominates over the magnetic one during almost the whole inflation stage, at the end of inflation it rapidly decreases and becomes subdominant. This can also be seen from the evolution of spectra in Fig.~\ref{fig-k-h} (c), (f), and (i). The electric energy density is transferred to charged fermions produced by the Schwinger effect. Indeed, the green curve corresponding to charged particles looks like a continuation of the blue curve representing the electric energy density.

Already from Figs.~\ref{fig-b10-SE}(a) and \ref{fig-b25-SE}(a) we can notice that $\tfrac{1}{2}|\langle\bm{E}\cdot\bm{B}\rangle|\approx \sqrt{\rho_{E}\rho_{B}}$ which is a signature of the fact that the electric and magnetic fields are nearly collinear. In order to check this more precisely, we plot the cosine of the angle between the electric and magnetic fields in the bottom plots in  Figs.~\ref{fig-b10-SE}(b) and \ref{fig-b25-SE}(b). They show that during the final part of inflation when the most significant generation of gauge fields occurs the cosine is greater than 0.8 and approaches unity at the end of inflation. This means that the angle between $\bm{E}$ and $\bm{B}$ is typically smaller than $35^{o}$ and decreases in time. This justifies our use of Eq.~(\ref{sigma}) for the Schwinger conductivity which was derived in the case of collinear electric and magnetic fields.

\section{Conclusion}
\label{sec-concl}

The explosive production of gauge fields during axion inflation is a complicated nonperturbative process. Its theoretical description becomes particularly challenging in the presence of nonlinear effects such as (i) the backreaction of the produced gauge fields on the evolution of the inflaton field and (ii) the Schwinger pair production of charged particles in the strong gauge-field background. In this paper, we have presented a novel gradient expansion formalism that is capable of successfully tackling this theoretical challenge, putting us in the position to qualitatively study and quantitatively describe the dynamics of gauge-field production during axion inflation at an unprecedented accuracy.

The typical approach to particle production during axion inflation, as it can be commonly found in the literature, consists in analyzing the gauge-field mode functions in Fourier space. This approach works well in the absence of backreaction or Schwinger pair production, in which case all Fourier modes evolve independently in time. However, as soon as at least one of these two nonlinear effects becomes relevant, the evolution of the individual Fourier modes can no longer be disentangled. All modes are coupled to each other, which complicates the theoretical and, in particular, numerical description of the system. In this case, backreaction can be accounted for by an iterative procedure that needs to be repeated until the numerical solutions for the gauge-field mode functions converge to a self-consistent result~\cite{Cheng:2015,Notari:2016,Domcke:2020}. The Schwinger effect, possibly in combination with backreaction, can in principle be treated in the same way, although to the best of our knowledge no such self-consistent iterative analysis in Fourier space has thus far appeared in the literature. In contrast to this approach in Fourier space, the main idea behind our gradient expansion formalism is to study the evolution of a set of bilinear electromagnetic functions in position space. These functions are defined in terms of scalar products of two electric or magnetic field vectors with an arbitrary power of the curl operator. Thus by construction, they automatically include all physically relevant gauge-field modes that experience a tachyonic instability and are hence excited above the vacuum level during axion inflation. This characteristic property is extremely helpful in the presence of mode coupling because of nonlinear effects such as backreaction and Schwinger pair production. In this case, our bilinear functions allow us to capture the dynamics of the entire system, without the need for explicitly disentangling the evolution of the individual Fourier modes.

In this paper, we derived and solved a set of evolution equations for the bilinear electromagnetic functions in the gradient expansion formalism. These equations are coupled into an infinite chain, which, however, can be truncated at some point based on well-justified physical arguments. During inflation, the number of modes that leave the horizon, undergo the quantum to classical transition and become tachyonically unstable constantly grows. This effect is accounted for in our evolution equations by nontrivial boundary terms. We derived new explicit expressions for them in terms of the Whittaker function and its derivative as well as accurate approximate expressions that are convenient for numerical applications. A novel feature of the boundary terms, which can be regarded as vacuum source terms in our evolution equations, is the fact that they are proportional to a new parameter $\Delta$. This parameter is related to the time integral over the time-dependent conductivity and describes the exponential damping of gauge fields deep inside the horizon caused by the presence of the conductive plasma. The state of the vacuum fluctuations inside the horizon thus depends on the evolution history at all earlier times, which renders the description of the system nonlocal in time and which led us to the notion of a time-dependent damped Bunch-Davies vacuum. As we were able to demonstrate, our gradient expansion formalism results in highly accurate and self-consistent solutions with a remaining numerical uncertainty of at most one to two percent. Moreover, it does not require an iterative procedure; all results can be generated in a single numerical integration of our differential equations.

While our formalism does not rely on any kind of spectral information in Fourier space, it can certainly be used as a starting point for solving the equations of motion for the gauge-field mode functions in Fourier space and hence derive approximate solutions for the energy spectra of the electric and magnetic fields. To this end, one only needs to use the numerical output of our formalism, specifically, the time evolution of the Hubble rate $H$, conductivity $\sigma$, and inflaton field $\phi$, as input in the gauge-field mode equation (\ref{A_1}). As shown in the previous section, this procedure results in highly accurate self-consistent energy spectra (see Fig.~\ref{fig-k-h}), which one can use for further phenomenological applications. This includes, e.g., an accurate calculation of the primordial scalar and tensor power spectra generated during axion inflation, notably, in the presence of nonlinear effects such as backreaction and Schwinger pair production. The first analysis along these lines, focusing on the scalar power spectrum and neglecting the Schwinger effect, has been carried out in Ref.~\cite{Domcke:2020}. An important outcome of this analysis was that backreaction can cause nontrivial features in the scalar power spectrum, which are related to the nontrivial evolution that we see in Fig.~\ref{fig-b20-25-noSE-rho-xi}. In this paper, on the other hand, we were able to show that, at fixed coupling strength $\beta$, backreaction effects can be suppressed because of efficient Schwinger pair production, see Fig.~\ref{fig-b25-SE}. An interesting open question therefore is whether backreaction effects ever have a chance to leave a noticeable imprint in the primordial scalar and tensor power spectra in the presence of efficient Schwinger pair production, if one is willing to consider significantly larger values of $\beta$. A numerical study of the strong-coupling regime, $\beta \gg \mathcal{O}(10)$, is, however, technically challenging, which is why we leave it for future work.

Another possible application of our formalism consists in a refined description of baryogenesis from hypermagnetic fields~\cite{Fujita:2016igl,Kamada:2016eeb,Kamada:2016cnb,Jimenez:2017cdr,Domcke:2019}. This baryogenesis mechanism relies on the observation that helical hypermagnetic fields generated in the early Universe, possibly during axion inflation, can source a primordial baryon asymmetry around the time of the electroweak phase transition by virtue of the chiral anomaly of baryon number in the Standard Model (see also Ref.~\cite{Domcke:2020quw} for an extended scenario, which in addition involves right-handed neutrinos). The most important input quantity for these baryogenesis scenarios is the amount of hypermagnetic helicity generated during primordial magnetogenesis. Based on our formalism, this quantity can now be accurately calculated as a function of the model parameters of axion inflation and in the presence of nonlinear effects. Finally, our formalism is of course useful with regard to the phenomenology of primordial magnetic fields on their own, which we briefly discussed in the Introduction. For a given model of axion inflation, our formalism can be used to determine the electric and magnetic power spectra towards the end of inflation, which can then serve as the starting point for studying the further evolution of these fields during reheating and beyond. In particular, our formalism can generate input spectra for magnetohydrodynamics simulations in the radiation-dominated era, if the impact of reheating on our spectra is assumed to be small. 

In the numerical analysis in this paper, we mostly considered a particular toy model in which axion inflation towards the end of inflation can be described by a simple quadratic mass term around the origin in field space. In the next step, it will be interesting to extend this analysis beyond this simple model and study the generation of the electric and magnetic fields during axion inflation in a model-independent way. In the literature, this is typically done in terms of two effective parameters: the inflationary Hubble rate $H$ and the parameter $\xi$, which quantifies the inflaton velocity during inflation in Hubble units. An important lesson from our analysis, however, is that such a description will no longer be possible in the presence of the Schwinger effect. In this case, one will need to consider the dependence on at least one more parameter, namely, the parameter $\Delta$, which describes the damping of the Bunch-Davies vacuum inside the Hubble horizon. We will return to this important question in future work. 

\begin{acknowledgments}
	We would like to thank Anastasiia Lysenko for her participation at the initial stage of the work as well as Azadeh Maleknejad for useful discussions. O.~O.~S.\ is grateful to the CERN Theory Group, where part of this work was done, for its kind hospitality.
	The work of E.~V.~G. was supported by the National Research Foundation of Ukraine Project No.~2020.02/0062.	
	This project has received funding from the European Union's Horizon 2020 Research and Innovation Programme under Grant Agreement No.~796961, ``AxiBAU'' (K.~S.).
	The work of O.~O.~S. was supported by the ERC-AdG-2015 Grant No.~694896 and the Swiss National Science Foundation Excellence Grant No.~200020B\_182864.
	The work of S.~I.~V. was supported by the Swiss National Science Foundation Grant No.~SCOPE IZSEZ0 206908.
\end{acknowledgments}

\appendix

\section{Schwinger conductivity in the Standard Model}
\label{app-particles}

The contribution of fermions and scalars to the Schwinger conductivity is given by Eqs.~(\ref{sigma}) and (\ref{sigma-scalar}), respectively. If there are several fermionic and scalar species, then the conductivity takes the form
\begin{multline}
\label{sigma-general}
    \sigma_{\rm full}=\frac{e^{3}}{12\pi^{2}}\frac{\sqrt{\mathcal{B}^{(0)}}}{H}\Bigg[\sum_{\mathrm{f}}N_{\mathrm{f}}|Q_{\mathrm{f}}|^{3}{\rm coth}\Big(\pi\sqrt{\frac{\mathcal{B}^{(0)}}{\mathcal{E}^{(0)}}}\Big)\exp\Big(-\frac{\pi m_{\mathrm{f}}^{2}}{|eQ_{\mathrm{f}}|\sqrt{\mathcal{E}^{(0)}}}\Big)+
    \\   +\sum_{\mathrm{s}}N_{\mathrm{s}}|Q_{\mathrm{s}}|^{3}{\rm cosech}\Big(\pi\sqrt{\frac{\mathcal{B}^{(0)}}{\mathcal{E}^{(0)}}}\Big)\exp\Big(-\frac{\pi m_{\mathrm{s}}^{2}}{|eQ_{\mathrm{s}}|\sqrt{\mathcal{E}^{(0)}}}\Big)\Bigg],
\end{multline}
where $N_{\mathrm{f}}$ and $N_{\mathrm{s}}$ are numbers of degrees of freedom (including spin) for fermions and scalars with charges $Q_{\mathrm{f}}$ and $Q_{\mathrm{s}}$ and masses $m_{\mathrm{f}}$ and $m_{\mathrm{s}}$.

As we discussed in Sec.~\ref{sec-model}, we consider the hypercharge $U(1)_{Y}$ group of the Standard Model in the symmetric phase, therefore, $e=g^{\prime}$, $Q=Y$. All fermions are massless, $m_{\mathrm{f}}=0$, while the mass of the Higgs field $m_{\mathrm{H}}$ is very large which renders the Higgs vacuum expectation value close to zero during inflation. We assume that $m_{\mathrm{H}}\gg g'\sqrt{\mathcal{E}^{(0)}}$ so that the Higgs field does not contribute to the Schwinger conductivity. Then Eq.~(\ref{sigma-general}) reads as
\begin{equation}
    \sigma_{\mathrm{SM}}=\frac{g^{\prime 3}}{12\pi^{2}}\sum_{\mathrm{f}}N_{\mathrm{f}}|Y_{\mathrm{f}}|^{3}\frac{\sqrt{\mathcal{B}^{(0)}}}{H}{\rm coth}\Big(\pi\sqrt{\frac{\mathcal{B}^{(0)}}{\mathcal{E}^{(0)}}}\Big).
\end{equation}

In Table~\ref{tab-SM} we list all SM fermions with their hypercharges and calculate their contributions to the Schwinger conductivity. Using this table, we obtain Eq.~(\ref{sigma-SM}).

\begin{table}[!h]
	\begin{tabular}{c|c|c|c}
		\hline\hline
		Particle & $N_{\mathrm{f}}$ &  $Y_{\mathrm{f}}$
		& $N_{\mathrm{f}}|Y_{\mathrm{f}}|^{3}$ \\
		\hline
		$e_{R}$ & $1\times 3_{\rm gen}=3$ & $-1$ & $3$
		\\ 
		 $l_{L}=\left(\begin{array}{c}
		     \nu_{L}\\
		     e_{L}
		\end{array}\right)$ &  $2\times 3_{\rm gen}=6$  &  $-\frac{1}{2}$  & $\frac{3}{4}$ 
		\\ 
		 $u_{R}$ &  $1\times3_{\rm c}\times 3_{\rm gen}=9$  &  $\frac{2}{3}$  & $\frac{8}{3}$ 
		\\ 
		 $d_{R}$ &  $1\times3_{\rm c}\times 3_{\rm gen}=9$  &  $-\frac{1}{3}$  & $\frac{1}{3}$ 
		\\
         $q_{L}=\left(\begin{array}{c}
		     u_{L}\\
		     d_{L}
		\end{array}\right)$ &  $2\times3_{\rm c}\times 3_{\rm gen}=18$  &  $\frac{1}{6}$  &  $\frac{1}{12}$ 
		\\ \hline
         \multicolumn{3}{c|}{\textbf{Sum over all fermions:}}  & $\boldsymbol{\frac{41}{6}}$ 
		\\\hline\hline
	\end{tabular}
\caption{\label{tab-SM} The SM fermions and their hypercharges.}
\end{table}

\section{Solutions of the mode equation and asymptotic expressions for the boundary terms}
\label{app-Whittaker}

The differential equation
\begin{equation}
\label{Whittaker-eq}
    \frac{d^{2}w}{dy^{2}}+\left(-\frac{1}{4}+\frac{\kappa}{y}+\frac{1/4-\mu^{2}}{y^{2}}\right)w=0
\end{equation}
is known as the Whittaker equation and has two linearly-independent solutions, $M_{\kappa,\mu}$ and $W_{\kappa,\mu}$. For the purposes of this paper, however, only the function $W_{\kappa,\mu}$ is relevant. It can be expressed in terms of the Tricomi confluent hypergeometric function $U$ as follows:
\begin{equation}
	W_{\kappa,\mu}(y)=e^{-y/2}y^{\mu+1/2}U(\mu-\kappa+1/2;1+2\mu;y).
\end{equation}

Using 
Eq.~(13.5.2) in Ref.~\cite{AbramowitzStegun}, one can derive the following asymptotical expression of the Whittaker function at $|y|\to \infty$,
\begin{equation}
\label{W-asym}
    W_{\kappa,\mu}(y)=e^{-y/2} y^{\kappa} [1+O(y^{-1})].
\end{equation}

The mode equation (\ref{A_3}) has the form of the Whittaker equation (\ref{Whittaker-eq}) with $\kappa=-i\lambda \xi$, $\mu=1/2+s$, and $y=2iz$. Its solution must satisfy the Bunch-Davies vacuum boundary condition 
(\ref{BD}). Comparing it with
Eq.~(\ref{W-asym}), we conclude that the $W$ function indeed has the correct asymptote. Therefore, the solution to the mode equation has form (\ref{A_Whittaker}).

In the derivation of boundary terms we used the expression for the derivative of the Whittaker function $W$ which is given by \{see Eq.~(13.4.33) in Ref.~\cite{AbramowitzStegun}\}
\begin{equation}
\label{Whittaker-derivative}
	y\frac{d}{dy}W_{\kappa,\mu}(y)=(y/2-\kappa)W_{\kappa,\mu}(y)-W_{\kappa+1,\mu}(y).
\end{equation}

Alternatively, the solution to Eq.~(\ref{A_3}) satisfying the boundary condition (\ref{BD}) can be expressed in terms of the Coulomb wave functions (see Sec.~14 in Ref.~\cite{AbramowitzStegun})
\begin{equation}
\label{A_Coulomb}
	f_{\lambda}(z,k)=\frac{1}{\sqrt{2k}}\left[G_{s}(\lambda \xi, -z) +iF_{s}(\lambda \xi, -z) \right].
\end{equation}
The boundary terms can also be expressed in terms of these functions as follows:
\begin{equation}
\label{E-Coulomb}
    E_{\lambda}(\xi,s)=G^{\prime 2}_{s}+F^{\prime 2}_{s}+\frac{2s}{r}\big[G_{s}G^{\prime}_{s}+F_{s}F^{\prime}_{s}\big]+\frac{s^{2}}{r^{2}}\big[ G^{2}_{s}+F^{2}_{s}\big],
\end{equation}
\begin{equation}
\label{G-Coulomb}
    G_{\lambda}(\xi,s)=-\big[G_{s}G^{\prime}_{s}+F_{s}F^{\prime}_{s}\big]-\frac{s}{r}\big[G^{2}_{s}+F^{2}_{s}\big],
\end{equation}
\begin{equation}
\label{B-Coulomb}
    B_{\lambda}(\xi,s)=G^{2}_{s}+F^{2}_{s},
\end{equation}
where $F_{s}=F_{s}(\lambda\xi,\,r(\xi,s))$ and all other functions have the same arguments; $r(\xi,s)$ is defined after Eq.~(\ref{B-lambda}) and prime denotes the derivative with respect to $r$, i.e., $F^{\prime}_{s}=\partial F_{s}(\lambda\xi,\,r)/\partial r$.

The functions $E_{\lambda}$, $G_{\lambda}$, and $B_{\lambda}$ which determine boundary terms are expressed in terms of the Whittaker functions and are given by Eqs.~(\ref{E-lambda})--(\ref{B-lambda}) [or by Eqs.~(\ref{E-Coulomb})--(\ref{B-Coulomb}) in terms of the Coulomb wave functions]. These expressions, however, are inconvenient for numerical computations for big values of $\xi$ because they require a significant increase of the working precision. Therefore, we found approximate expressions for these functions in terms of elementary functions which provide the sufficient accuracy for large values of $|\xi|$ and all relevant values of parameter $s$.

We start with the expressions for $s=0$. In such a case, there exist power series for the Whittaker functions (or the Coulomb wave functions) in inverse powers of $|\xi|$. For $\lambda={\rm sign\,}\xi$, it is convenient to work with the Coulomb wave functions. The asymptotical expansions for $F_{0}(|\xi|,\,2|\xi|)$, $G_{0}(|\xi|,\,2|\xi|)$, and the corresponding derivatives can be found from the integral representations of the Coulomb wave functions by the method discussed in Ref.~\cite{Abramowitz:1954} --- for a few first terms in these expansions, see Eqs.~(14.5.10) and (14.5.11) in Ref.~\cite{AbramowitzStegun} and also Refs.~\cite{Froeberg:1955,Isacson:1968}. Applying this method and using Eqs.~(\ref{E-Coulomb})--(\ref{B-Coulomb}), we finally get the expansions which reproduce the exact result with a relative error less than $10^{-4}\%$ for $|\xi|\geq 3$,
\begin{multline}
    E_{{\rm sign\,}\xi}(\xi,\,0)= \frac{(\tfrac{3}{2})^{1/3}\Gamma^{2}(\tfrac{2}{3})}{\pi\, |\xi|^{1/3}}-\frac{\sqrt{3}}{15\,|\xi|}+\frac{(\tfrac{2}{3})^{1/3}\Gamma^{2}(\tfrac{1}{3})}{100\,\pi\, |\xi|^{5/3}}+\frac{(\tfrac{3}{2})^{1/3}\Gamma^{2}(\tfrac{2}{3})}{1\,575\,\pi\, |\xi|^{7/3}}-\frac{27\sqrt{3}}{19\,250\,|\xi|^{3}}+\frac{359\, (\tfrac{2}{3})^{1/3}\Gamma^{2}(\tfrac{1}{3})}{866\,250\,\pi\, |\xi|^{11/3}}\\+\frac{8\,209\,(\tfrac{3}{2})^{1/3}\Gamma^{2}(\tfrac{2}{3})}{13\,162\,500\,\pi\, |\xi|^{13/3}}-\frac{690\,978\sqrt{3}}{1\,861\,234\,375\,|\xi|^{5}}+\frac{13\,943\,074\, (\tfrac{2}{3})^{1/3}\Gamma^{2}(\tfrac{1}{3})}{127\,566\,140\,625\, \pi\, |\xi|^{17/3}}+O(|\xi|^{-19/3}),
\end{multline}
\begin{multline}
    G_{{\rm sign\,}\xi}(\xi,\,0)=\frac{1}{\sqrt{3}}-\frac{(\tfrac{2}{3})^{1/3}\Gamma^{2}(\tfrac{1}{3})}{10\,\pi\, |\xi|^{2/3}}+ \frac{3(\tfrac{3}{2})^{1/3}\Gamma^{2}(\tfrac{2}{3})}{35\,\pi\, |\xi|^{4/3}}-\frac{\sqrt{3}}{175\,|\xi|^{2}}-\frac{41(\tfrac{2}{3})^{1/3}\Gamma^{2}(\tfrac{1}{3})}{34\,650\,\pi\, |\xi|^{8/3}}+\frac{10\,201(\tfrac{3}{2})^{1/3}\Gamma^{2}(\tfrac{2}{3})}{2\,388\,750\,\pi\, |\xi|^{10/3}}-\frac{8\,787\sqrt{3}}{21\,896\,875\,|\xi|^{4}}\\-\frac{1\,927\,529\, (\tfrac{2}{3})^{1/3}\Gamma^{2}(\tfrac{1}{3})}{4\,638\,768\,750\,\pi\, |\xi|^{14/3}}+\frac{585\,443\,081\,(\tfrac{3}{2})^{1/3}\Gamma^{2}(\tfrac{2}{3})}{393\,158\,390\,625\,\pi\, |\xi|^{16/3}}-\frac{65\,977\,497\sqrt{3}}{495\,088\,343\,750\,|\xi|^{6}}+O(|\xi|^{-20/3}),
\end{multline}
\begin{multline}
    B_{{\rm sign\,}\xi}(\xi,\,0)=\frac{(\tfrac{2}{3})^{1/3}\Gamma^{2}(\tfrac{1}{3})|\xi|^{1/3}}{\pi} +\frac{2\sqrt{3}}{35\,|\xi|}-\frac{4(\tfrac{2}{3})^{1/3}\Gamma^{2}(\tfrac{1}{3})}{225\,\pi\, |\xi|^{5/3}}+\frac{9(\tfrac{3}{2})^{1/3}\Gamma^{2}(\tfrac{2}{3})}{1\,225\,\pi\, |\xi|^{7/3}}+\frac{132\sqrt{3}}{56\,875\,|\xi|^{3}}-\frac{9\,511\, (\tfrac{2}{3})^{1/3}\Gamma^{2}(\tfrac{1}{3})}{5\,457\,375\,\pi\, |\xi|^{11/3}}\\
    +\frac{1\,448\,(\tfrac{3}{2})^{1/3}\Gamma^{2}(\tfrac{2}{3})}{1\,990\,625\,\pi\, |\xi|^{13/3}}+\frac{1\,187\,163\sqrt{3}}{1\,323\,765\,625\,|\xi|^{5}}-\frac{22\,862\,986\, (\tfrac{2}{3})^{1/3}\Gamma^{2}(\tfrac{1}{3})}{28\,465\,171\,875\,\pi\, |\xi|^{17/3}}+O(|\xi|^{-19/3}).
\end{multline}

For $\lambda=-{\rm sign\,}\xi$, asymptotical expansions for the Whittaker functions can be obtained by the method discussed in Sec.~6.13.3 of Ref.~\cite{Bateman_vol1} and Sec.~10 of Ref.~\cite{Erdelyi:1957}. Approximate expressions for the functions $E_{\lambda}$, $G_{\lambda}$, and $B_{\lambda}$ with an error less than $10^{-4}\%$ for $|\xi|\geq 3$ have the form
\begin{equation}
    E_{-{\rm sign\,}\xi}(\xi,\,0)=\sqrt{2} \Bigg[1-\frac{9}{2^{10} \xi^{2}}+\frac{2\,059}{2^{21} \xi^{4}}-\frac{448\,157}{2^{31} \xi^{6}}+O(|\xi|^{-8})\Bigg],
\end{equation}
\begin{equation}
    G_{-{\rm sign\,}\xi}(\xi,\,0)=-\frac{\sqrt{2}}{32|\xi|} \Bigg[1-\frac{67}{2^{10} \xi^{2}}+\frac{21\,543}{2^{21} \xi^{4}}-\frac{6\,003\,491}{2^{31} \xi^{6}}+O(|\xi|^{-8})\Bigg],
\end{equation}
\begin{equation}
    B_{-{\rm sign\,}\xi}(\xi,\,0)=\frac{1}{\sqrt{2}} \Bigg[1+\frac{11}{2^{10} \xi^{2}}-\frac{2\,397}{2^{21} \xi^{4}}+\frac{508\,063}{2^{31} \xi^{6}}+O(|\xi|^{-8})\Bigg].
\end{equation}

For $s\neq 0$, it is much more difficult to get similar expressions because the second parameter appears in the expansion. We used the results of Ref.~\cite{Biedenharn:1955} and modified the first few terms of our previous 
expansions for $s=0$ in order to obtain the approximate expressions reproducing exact results with an error of less than $0.5\%$ for $|\xi|\geq 4$ and all values of the parameter $s$. For $\lambda={\rm sign\,}\xi$, 
we obtain
\begin{multline}
    E_{{\rm sign\,}\xi}(\xi,\,s)\approx \frac{\psi^{1/3}}{r^{1/3}}\Bigg[ \frac{3^{1/3}\Gamma^{2}(\tfrac{2}{3})}{\pi}-\frac{2}{5\sqrt{3}}\Big(\frac{\psi}{r^{2}}\Big)^{1/3}\Big(1+\frac{5s}{\psi^{2/3}}\Big)+\frac{\Gamma^{2}(\tfrac{1}{3})}{3^{1/3}25\pi } \Big(\frac{\psi}{r^{2}}\Big)^{2/3} \Big(1+\frac{5s}{\psi^{2/3}}\Big)^{2}+\\
    +\frac{3^{1/3}4\Gamma^{2}(\tfrac{2}{3})}{1575\pi } \Big(\frac{\psi}{r^{2}}\Big) \Big(1-\frac{135s}{\psi^{2/3}}\Big)+\frac{4\sqrt{3}}{875}\Big(\frac{\psi}{r^{2}}\Big)^{4/3} \Big(-\frac{27}{11}+\frac{10s}{\psi^{2/3}}+\frac{25s^{2}}{\psi^{4/3}}\Big)\Bigg],
\end{multline}
\begin{equation}
    G_{{\rm sign\,}\xi}(\xi,\,s)\approx \frac{1}{\sqrt{3}}-\frac{\Gamma^{2}(\tfrac{1}{3})}{3^{1/3}5\pi }\Big(\frac{\psi}{r^{2}}\Big)^{1/3}\Big(1+\frac{5s}{\psi^{2/3}}\Big)+\frac{3^{1/3}6\Gamma^{2}(\tfrac{2}{3})}{35\pi }\Big(\frac{\psi}{r^{2}}\Big)^{2/3}-\frac{4\sqrt{3}}{175}\Big(\frac{\psi}{r^{2}}\Big)\Big(1+\frac{5s}{\psi^{2/3}}\Big),
\end{equation}
\begin{equation}
    B_{{\rm sign\,}\xi}(\xi,\,s)\approx\frac{r^{1/3}}{\psi^{1/3}}\Bigg[ \frac{\Gamma^{2}(\tfrac{1}{3})}{3^{1/3}\pi}+\frac{4\sqrt{3}}{35}\Big(\frac{\psi}{r^{2}}\Big)^{2/3}-\frac{16\Gamma^{2}(\tfrac{1}{3})}{3^{1/3}225\pi}\Big(\frac{\psi}{r^{2}}\Big)\Bigg],
\end{equation}
while for $\lambda=-{\rm sign\,}\xi$, we have
\begin{equation}
    E_{-{\rm sign\,}\xi}(\xi,\,s)\approx 2\left(\frac{|\xi|}{r}\right)^{1/2}\left(1+\frac{s}{16\xi^{2}}\frac{3|\xi|-r}{r}+\frac{s^{2}}{4|\xi|r}\right),
\end{equation}
\begin{equation}
    G_{-{\rm sign\,}\xi}(\xi,\,s)\approx -\frac{1}{16\sqrt{|\xi| r}}\left(\frac{3|\xi|-r}{|\xi|}+8s\right),
\end{equation}
\begin{equation}
    B_{-{\rm sign\,}\xi}(\xi,\,s)\approx \frac{1}{2}\left(\frac{r}{|\xi|}\right)^{1/2},
\end{equation}
where
\begin{equation}
r=r(\xi,\,s)\equiv |\xi|+\sqrt{\xi^{2}+s^{2}+s}, \qquad \psi=\psi(\xi,\,s)\equiv \frac{2\sqrt{\xi^{2}+s^{2}+s}}{|\xi|+\sqrt{\xi^{2}+s^{2}+s}}.
\end{equation}

 \end{document}